\documentclass[10pt,letterpaper]{article}
\usepackage[top=0.85in,left=2.75in,footskip=0.75in]{geometry}

\usepackage{changepage}
\usepackage{textcomp,marvosym}
\usepackage{fixltx2e}
\usepackage{amsmath,amssymb}
\usepackage{nameref,hyperref}
\usepackage[utf8]{inputenc}
\usepackage{microtype}
\DisableLigatures[f]{encoding = *, family = * }
\usepackage{rotating}

\usepackage[backend=bibtex]{biblatex}
\bibliography{template-1}

\raggedright
\usepackage[aboveskip=1pt,labelfont=bf,labelsep=period,justification=raggedright,singlelinecheck=off]{caption}

\date{}

\begin{document}
\vspace*{0.35in}

\begin{flushleft}
{\Large
\textbf\newline{Random wiring, ganglion cell mosaics, and the functional architecture of the visual cortex}
}
\newline
\\
Manuel Schottdorf\textsuperscript{1-5,\Yinyang},
Wolfgang Keil\textsuperscript{1-4,5,\Yinyang,*},
David Coppola\textsuperscript{7},
Leonard E. White\textsuperscript{8},
Fred Wolf\textsuperscript{1-4,9}
\\
\bigskip
\bf{1} Max Planck Institute for Dynamics and Self-organization, 37077 G\"ottingen, Germany
\\
\bf{2} Bernstein Center for Computational Neuroscience, 37077 G\"ottingen, Germany
\\
\bf{3} Bernstein Focus for Neurotechnology, 37077 G\"ottingen, Germany
\\
\bf{4} Faculty of Physics, University of G\"ottingen, 37077 G\"ottingen, Germany
\\
\bf{5} Institute for Theoretical Physics, University of W\"urzburg, 97074, W\"urzburg, Germany
\\
\bf{6} Center for Studies in Physics and Biology, The Rockefeller University, New York, NY, 10065, USA
\\
\bf{7} Department of Biology, Randolph-Macon College, Ashland, VA 23005, USA
\\
\bf{8} Department of Orthopaedic Surgery, Duke Institute for Brain Sciences, Duke University, Durham NC 27708, USA
\\
\bf{9} Kavli Institute for Theoretical Physics, Santa Barbara, CA 93106-4030, USA.
\bigskip

%
%
\Yinyang These authors contributed equally to this work.

* wkeil@rockefeller.edu

\end{flushleft}
\section*{Abstract}
The architecture of iso-orientation domains in the primary visual cortex (V1) of placental carnivores and primates apparently follows species invariant quantitative laws. Dynamical optimization models assuming that neurons coordinate their stimulus preferences throughout cortical circuits linking millions of cells specifically predict these invariants. This might indicate that V1's intrinsic connectome and its functional architecture adhere to a single optimization principle with high precision and robustness. To validate this hypothesis, it is critical to closely examine the quantitative predictions of alternative candidate theories. Random feedforward wiring within the retino-cortical pathway represents a conceptually appealing alternative to dynamical circuit optimization because random dimension-expanding projections are believed to generically exhibit computationally favorable properties for stimulus representations.\\
Here, we ask whether the quantitative invariants of V1 architecture can be explained as a generic emergent property of random wiring. We generalize and examine the stochastic wiring model proposed by Ringach and coworkers, in which iso-orientation domains in the visual cortex arise through random feedforward connections between semi-regular mosaics of retinal ganglion cells (RGCs) and visual cortical neurons. We derive closed-form expressions for cortical receptive fields and domain layouts predicted by the model for perfectly hexagonal RGC mosaics. Including spatial disorder in the RGC positions considerably changes the domain layout properties as a function of disorder parameters such as position scatter and its correlations across the retina. However, independent of parameter choice, we find that the model predictions substantially deviate from the layout laws of iso-orientation domains observed experimentally. Considering random wiring with the currently most realistic model of RGC mosaic layouts, a 
pairwise interacting point process, the predicted layouts remain distinct from experimental observations and resemble Gaussian random fields. We conclude that V1 layout invariants are specific quantitative signatures of visual cortical optimization, which cannot be explained by generic random feedforward-wiring models.
%
\section*{Author Summary}
In the primary visual cortex of primates and carnivores, local visual stimulus features such as edge orientation are processed by neurons arranged in arrays of iso-orientation domains. Large-scale comparative studies have uncovered that the spatial layout of these domains and their topological defects follows species-invariant quantitative laws, predicted by models of large-scale circuit self-organization. Here, we ask whether the experimentally observed layout invariants might alternatively emerge as a consequence of random connectivity rules for feedforward projections from a small number of retinal cells to a much larger number of cortical target neurons. In this random wiring framework, the semi-regular and spatially granular arrangement of retinal ganglion cells determines the spatial layout of visual cortical iso-orientation domains -- a hypothesis diametrically opposed to cortical large-scale circuit self-organization. Generalizing a prominent model of the early visual pathway, we find that the random 
wiring framework does not reproduce the experimentally determined layout invariants. Our results demonstrate how comparison between theory and quantitative phenomenological laws obtained from large-scale experimental data can successfully discriminate between competing hypotheses about the design principles of cortical circuits.


\section*{Introduction}
\begin{figure}
\begin{center}
\includegraphics[width=\linewidth]{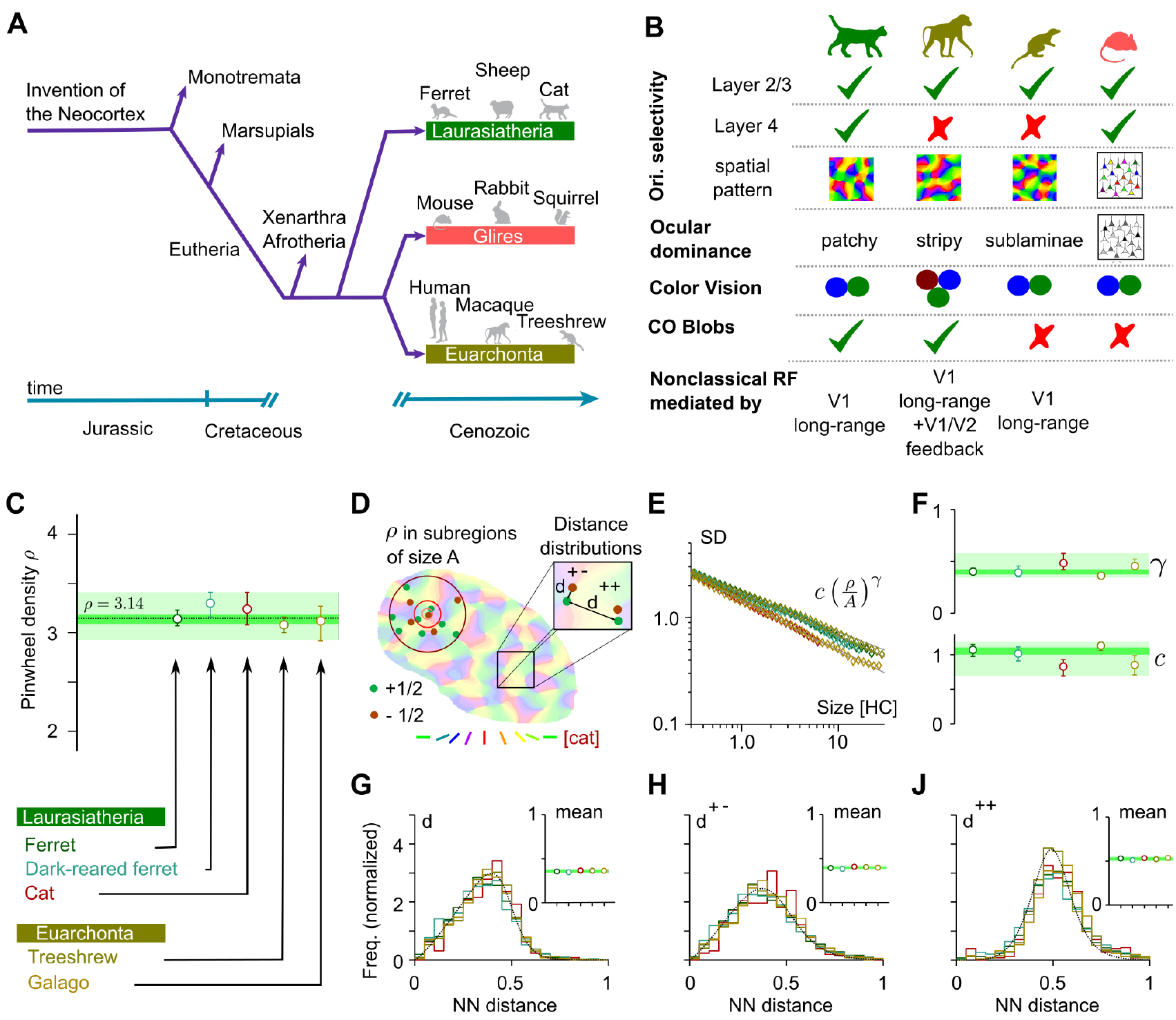}
\end{center}
\caption{\textbf{Common laws for the layout of iso-orientation domains in different mammalian species.}
\textbf{A} Phylogenetic relationships and macroevolution of laurasiatheria, euarchonta and glires \cite{Murphy2001, Kaas2008, Kriegs2006, Keil2012, Kaas2012, Kaschube2014}.
\textbf{B} Key features of the thalamo-cortical pathway for cat \cite{Alonso2001, Usrey1999, Ferster1996, Reid1995, Thompson1994, Shou1989, Levick1980, Levay1976, Hammond1974, Hubel1962, Boycott1974, Hubel1963, Hubel1961, Hubel1959, Ohki2006}, macaque \cite{Blasdel1984, Ringach2002, Gur2005, Smith1990, Essen1978, Schall1986, Obermayer1993, Irvin1993}, treeshrew \cite{Humphrey1977, Kretz1986, Fitzpatrick1996, Bosking1997, Chisum2003, VanHooser2013} and mouse \cite{Drager1975, Ohki2005, Niell2008, Elstrott2008,Marshel2012} at the level of retina/LGN and layer IV and II/III of V1. All species show orientation selective neurons in layer II/III, but only cat, ferret, and mouse exhibit orientation selectivity in input layer IV. Ocular dominance domain layouts differs greatly between all four species, macaque is the only species listed possessing trichromatic color vision. Only cat and macaque V1 display cytochrome oxidase blobs.  Non-classical receptive fields are mediated by different circuits in cat, tree shrew 
and macaque.
\textbf{C} Pinwheel density $\rho$ in ferret (N = 82), dark-reared ferret (N = 21), cat (N = 13), tree shrew (N = 26), and galago (N = 9). Light green shading indicates one--species consistency range, dark green shading indicates common design consistency range (see text).
\textbf{D} Illustration of the common design layout features, nearest neighbor (NN) distances, and pinwheel density in subregions of varying size.
\textbf{E, F} Standard deviations ($SD$) of pinwheel densities as a function of the area A of randomly selected subregions. $SD(A)$ is well described by a power law with variability exponent $\gamma$ (\textbf{F}, top) and variability coefficient $c$ (\textbf{F}, bottom). 
\textbf{(G-J)} Nearest neighbor distances distributions for pinwheels of arbitrary (G), opposite (H) and equal (J) topological charge in units of the column spacing. Insets indicate species means. All error bars represent 95\% confidence intervals of the bootstrap distributions.}
\label{fig_0}
\end{figure}

Processing high-dimensional external stimuli and efficiently communicating their essential features to higher brain areas is a fundamental function of any sensory system.  For many sensory modalities, this task is implemented via convergent and divergent neural pathways in which information from a large number of sensors is compressed into a smaller layer of neurons, transmitted, and then re-expanded into a larger neuronal layer. When sensory inputs are sparse, compression of the inputs through random convergent feedforward projections has been shown to retain much of the information present in the stimuli \cite{Bruckstein2009,Ganguli2010, Ganguli2012}. On the other hand, random expanding projections can lead to computationally powerful high-dimensional representations of such compressed signals, which combine separability of the inputs with high signal-to-noise ratio to facilitate downstream readouts \cite{Babadi2014}. Given these computational benefits, one might expect randomness to be a fundamental 
wiring principle employed by different sensory systems. The most striking example of a random expansion so far has been observed in the olfactory system of \textit{Drosophila melanogaster}. Kenyon cells in the fly brain's mushroom body were shown to integrate input from various olfactory glomeruli in combinations that are consistent with purely random choices from the overall distribution of glomerular projections to the mushroom body \cite{Caron2013}.\\
What is the role of random projections between neural layers in mammalian sensory systems? Sompolinsky and others have argued that the human visual system, for instance, implements a compression-transmission-expansion strategy \cite{Ganguli2012, Babadi2014}. In fact, visual stimulus information is transmitted from about 5 million cone photoreceptors \cite{Jonas1992,Curcio1990} to 1 million retinal ganglion cells (RGCs) \cite{Curcio1990} and then via the optic nerve to about 1 million lateral geniculate relay cells \cite{Selemon2007} to on the order of 100 million neurons in the primary visual cortex (V1) \cite{Klekamp1991, Leuba1994}. We note that, while the overall connectivity indeed suggests compression for peripheral retinal regions \cite{Field2010}, close to the fovea RGC density is higher than the density of photoreceptors \cite{Wassle1989, Wassle1990}.\\
How much does randomness contribute to shaping the functional architecture of early visual cortical areas? Projections between individual layers of the early mammalian visual pathway are clearly not entirely random. Visual information is mapped visuotopically from the retina to V1 such that neighboring groups of V1 neurons process information from neighboring regions in visual space. Yet, it has long been realized that many features of the spatial progression of receptive fields across V1 layer IV naturally emerge if random feedforward connections from groups of RGCs cells to layer IV neurons (via the lateral geniculate nucleus (LGN)) are assumed (see \cite{Malsburg1973} for an early example). The most important of such features is orientation selectivity, i.e. the selective response to edge-like stimuli of a particular orientation. In carnivores, primates and their close relatives, orientation selectivity is arranged in patterns of iso-orientation domains. Iso-orientation domains (orientation domains for 
short) in V1 exhibit a continuous, roughly repetitive arrangement. A distance in the millimeter range, called the column spacing, separates neighboring domains preferring the same orientation. The continuous progression of preferred orientations is interrupted by a system of topological defects, called pinwheel centers, at which neurons selective to the whole complement of stimulus orientations are located in close vicinity \cite{Grinvald1986, Blasdel1986, Swindale1987, Bonhoeffer1991, Bosking1997, Chapman1996}. These topological defects exhibit two distinct topological charges, indicating that preferred orientations change clockwise or counterclockwise around the defect center \cite{Swindale1982, Grinvald1986, Bonhoeffer1991, Wolf1998, Kaschube2010}. \\
More than 25 years ago, Soodak \cite{Soodak1986,Soodak1987} (see also \cite{Wassle1981}) proposed random wiring between irregularly positioned retinal ganglion cells (RGCs) and layer IV neurons in V1 via the thalamus as a candidate mechanism defining the pattern of iso-orientation domains. According to this statistical wiring hypothesis, a V1 neuron randomly samples feedforward inputs from geniculate projections in the immediate vicinity of its receptive field center (see e.g. \cite{Alonso2001}). The neuron then is likely to receive the strongest inputs from a central pair of ON/OFF RGCs, forming a so-called RGC dipole \cite{Paik2011, Paik2012, Paik2014}. In this scheme, one ON and one OFF subregion dominate the receptive field (RF) of the V1 neuron and its response is tuned to the orientation perpendicular to the dipole axis. Thus, the preferred orientation of the neuron in this case is determined by the orientation of the RGC dipole. Consequently, the key prediction of the statistical wiring hypothesis is 
that the spatial arrangement of ON/OFF RGC cells in the retina essentially determines the spatial layout of orientation preference domains in V1. \\
Recently, Paik \& Ringach showed that the statistical wiring hypothesis -- when constructed with a hexagonal grid of RGCs -- predicts a periodic orientation domain layout with a hexagonal autocorrelation function \cite{Paik 2011}. Moreover, it predicts that orientation preference is differently linked to the visuotopic map around pinwheels of positive or negative topological charge \cite{Paik2012}. Qualitative signatures of both predictions were reported to be present in experimentally measured patterns \cite{Paik2011, Paik2012}. Thus the statistical wiring model has conceptual appeal and is a mechanistically particularly transparent candidate explanation for V1 functional architecture (see however \cite{Hore2012, Schottdorf2014}). Does the predictive power of the random wiring hypothesis for the early visual pathway reach beyond this qualitative agreement? \\
The recent discovery of species-invariant quantitative layout laws for the arrangement of pinwheel centers in tree shrews, galagos and ferrets \cite{Kaschube2010} provides a unique opportunity to address this question. Kaschube et al. demonstrated that in these species, the statistics of pinwheel defect layouts is quantitatively invariant, with potential deviations in geometrical layout parameters of at most a few percent \cite{Kaschube2010}. Specifically, the overall pinwheel density, defined as the average number of defects within the area of one square column spacing $\Lambda^2$ was found to be virtually identical. Subsequently, orientation domain layouts from cat V1 were shown to exhibit pinwheel densities very close to those of the three species previously studied \cite{Keil2012}. Additionally, Kaschube et al. found an entire set of local and non-local quantitative pinwheel layout features to be species-invariant (see below). Following \cite{Kaschube2010}, we refer to this overall layout of orientation 
domains as the \textit{common design}. \\
During mammalian evolution, the common design most likely arose independently in carnivores and euarchontans and potentially even in scandentia.\cite{Kaschube2010, Kaas2012}. This is suggested by two lines of evidence: (i) The four species in which the common design has been observed so far are widely separated in terms of evolutionary descent, belonging to distinct supra-ordinal clades that split already during basal radiation of placentals \cite{Murphy2001,Kriegs2006, Kriegs2007,BinindaEmonds2007, Meredith2011,OLeary2013, Springer2013, OLeary2013b} (\textbf{Fig.~\ref{fig_0}A}, see also \cite{Kaschube2010, Keil2012}). Their last common ancestor was a small shrew-like mammal \cite{OLeary2013, Springer2013, OLeary2013b} that is unlikely to have possessed a columnar V1 architecture \cite{Kaschube2010, Kaas2012}. (ii) Distinct neuronal circuits underlie the generation of orientation selectivity in galago, ferret, tree shrew, and cat (\textbf{Fig.~\ref{fig_0}B}). Tree shrews, for instance, lack orientation 
selectivity in the input layer IV of V1 \cite{Chisum2003, VanHooser2013} and use intracortical circuits to compute contour orientation. In contrast, cats exhibit both, orientation selectivity and organization of selectivity into orientation domains already in layer IV and thus first generate orientation selectivity by thalamo-cortical circuits \cite{Hubel1962, Hubel1963} (see \textbf{Fig.~\ref{fig_0}B} for further differences).\\
Kaschube et al. used a dynamical self-organization model with long-range suppressive interactions, the long-range interaction model, to explain all features of the common design \cite{Kaschube2010}. The hypothesis that randomness of feedforward connections between the retina/LGN and V1 could explain the common design is conceptually diametrically opposed to large-scale self-organization. In the long-range interaction model, the orientation preference of a neuron is chosen from an, in principle, unlimited afferent repertoire of potential receptive fields. Single neurons dynamically select a particular preferred orientation as a result of large-scale circuit interactions involving millions of other cortical neurons. In the statistical connectivity model, to the contrary, the preferred orientation of a cortical neuron is essentially imposed by the alignment of only one pair of neighboring ON-OFF RGCs, a local process involving in principle not more than 5 cells. 
Can the invariant layout laws of iso-orientation domains and pinwheels be explained as the generic outcome of a locally stochastic feedforward wiring of the early visual pathway? More generally, do iso-orientation domains and pinwheels in different species adhere to identical layout laws because any mechanism that generates a retinotopic random feedforward circuit will automatically set up a layout that adheres to the common design?\\	
Here, we systemically investigate the arrangements of iso-orientation domains generated by the statistical connectivity model and assess their consistency with the experimentally observed common design invariants. First, we consider the statistical wiring model with perfectly hexagonal mosaics of RGCs, its most tractable form.  We derive closed-form expressions for cortical neuron receptive fields and orientation domain layouts resulting from the Moir\'{e} interference effect of hexagonal ON and OFF ganglion cell mosaics \cite{Paik2011, Paik2012}. The pinwheel density of these pinwheel layouts is $\rho=2\sqrt{3}\approx3.46$, substantially larger than experimentally observed. We then characterize the orientation domain layouts resulting from spatially disordered hexagonal mosaics. We find that parameters of RGC position disorder can not be tuned such that the statistical wiring model's layouts match the quantitative invariants of the common design. Next, we examine a generalized class of noisy hexagonal 
mosaics that allows for spatially correlated disorder of RGC positions. This correlated retinal disorder induces local variations in column spacing, mimicking column spacing heterogeneity in the visual cortex \cite{Kaschube2002, Kaschube2009}. With these mosaics, Moir\'{e} interference persists to larger disorder strength. Pinwheel densities, however, are unaffected by low and intermediate levels of disorder and increase from a lower bound of 3.5 for stronger disorder. Finally, we characterize the statistical connectivity model with RGC mosaics generated by Eglen's pairwise interaction point process (PIPP), the most realistic model for RGC mosaics currently available \cite{Eglen2005, Hore2012, Schottdorf2014}. The resulting arrangements of iso-orientation domains and pinwheels are identical to those predicted by Gaussian random field models \cite{Wolf1998, Wolf2003, Schnabel2007}. Their pinwheel densities can be tuned by applying band pass filters of different bandwidths. However, for all plausible filter 
shapes, pinwheel densities are substantially larger than experimentally observed. \\
Our findings demonstrate that the mechanism for seeding patterns of iso-orientation domains described by the stochastic wiring model predicts column arrangements substantially different from the long-range interaction model and distinct from the experimentally observed invariant common design. %
%
%
%

\pagebreak
%
%
\section*{Results}
\subsection*{A benchmark for models of orientation domains in V1}
Our overall goal was to assess whether the layout of orientation domains predicted by the statistical wiring model are consistent with the observed common design invariants. To achieve this, we first sought to establish a benchmark for model of orientation domain layouts in general, to which predicted layouts can then be compared. To this end, we re-analyzed the data set used in \cite{Kaschube2010} using the fully automated method described in the same study. The data set contains optical imaging of intrinsic signal experiments from tree shrew  (N = 26), ferrets (N = 82), dark-reared ferrets  (N = 21) and galagos (N = 9). Because many previous studies used the statistical wiring model with parameters optimized to mimic the early visual pathway of the cat, e.g. \cite{Ringach2007}, we additionally analyzed data from 13 cat V1 hemispheres. \\
Following \cite{Kaschube2010}, we first computed the average pinwheel densities (\textbf{Fig.~\ref{fig_0}C}). Pinwheel densities of all four species, including cat were statistically indistinguishable from each other and statistically indistinguishable from $\pi$ (dark-reared ferrets excluded) - the value predicted for the average pinwheel density by the long-range interaction model \cite{Kaschube2010}.  
As a measure of pinwheel position variability, spanning all scales from single hypercolumn to the entire imaged region, we calculated the standard deviation, SD, of pinwheel density estimates in circular subregions of area $A$ (see \textbf{Fig.~\ref{fig_0}D} for an illustration). For all species, the function $\text{SD}(A)$ was well described by
\begin{eqnarray}
\text{SD}(A) = c\left(\frac{\rho}{A}\right)^{\gamma}\,\label{sdfun}
\end{eqnarray}
(\textbf{Fig.~\ref{fig_0}E}) with $\rho$ denoting the average pinwheel density. The variability exponents $\gamma$ and variability coefficients $c$ were similar in all four species (\textbf{Fig.~\ref{fig_0}F}). As a measure of relative pinwheel positioning on the hypercolumn scale, we  computed the nearest neighbor (NN) distance statistics for pinwheels of same or opposite topological charge as well as independent of their topological charge (see \textbf{Fig.~\ref{fig_0}D} for an illustration). Distance distributions were unimodal and very similar (\textbf{Fig.~\ref{fig_0}G-J}). Importantly, the distributions obtained from cat V1 were indistinguishable from the other three species. Mean NN distances, when measured in units of hypercolumns, were statistically indistinguishable (\textbf{Fig.~\ref{fig_0}G-J}, insets).
These findings confirm the results of \cite{Kaschube2010,Keil2012} and show that cat primary visual cortex follows the same quantitative layout laws as in tree shrew, galago and ferret. \\
From the above results, we extracted two types of consistency ranges that can be used as a benchmark for models of orientation domains in V1. To be consistent with an observed layout of orientation domains, a model's predictions should not be significantly different from experimental observations in at least one species. We thus defined one species consistency ranges spanned by the minimal lower and maximal upper margin of the single species confidence intervals for each parameter. If a model's predicted layout parameters are located outside one or more of the one species consistency ranges, data from every species rejects this model at 5\% significance level. This criterion is thus conservative in nature and does not assume that there is in fact one species invariant common design. If such a truly universal common design for orientation domains in fact exists, it would be appropriate to pool data from different species and to consider the more precisely defined confidence intervals of the grand average 
statistics as the relevant benchmark. To perform this more demanding test of model viability, we also defined common design consistency ranges as the 95\% bootstrap confidence intervals obtained from the whole data set. If the layout parameters predicted by a model are within all of the common design consistency ranges, the model offers a quantitative account of the bona fide universal common design. If one or more layout parameters have predicted values outside the common design consistency ranges the model is inconsistent with the common design. With the current data set, if a model is common design consistent, it is also one species consistent. One species (common design) consistency ranges are shaded in light (dark) green in \textbf{Fig.~\ref{fig_0}C,E-J}  and summarized in \textbf{Tab.~\ref{commdsgn}}. The tests of model viability defined above are most simply performed if the parameter values predicted by a model are determined exactly or with a numerical error that is much smaller than the empirical 
uncertainties. For models that can be solved accurately numerically, this can in principle always be achieved by a sufficiently large sample size of simulations. In the following, through analytical and numerical calculations, we will perform a comprehensive search through the statistical wiring model's parameter space to identify regimes in which the model is one species consistent or common design consistent. 
\begin{table}[t]
\begin{adjustwidth}{-1.5in}{0in} 
\begin{small}
    \begin{tabular}{ | p{3cm} || p{1.6cm} | p{1.85cm} | p{1.85cm} | p{1.85cm} | p{1.75cm}| p{1.75cm}|}
    \hline
     & Pinwheel density $\rho$ & NN distance ind. charge & NN distance same charge & NN distance opp. charge & Variab. exp. $\gamma$ & Variab. coeff. $c$ \\ \hline\hline
     Ferret & 3.14 & 0.355 & 0.523 & 0.393 & 0.40 & 1.07\\
      & [3.06, 3.23] & [0.347, 0.363] & [0.521, 0.539] & [0.383, 0.403] & [0.37, 0.44] & [0.97, 1.15]\\ \hline
     Dark-reared Ferret& 3.30 & 0.346& 0.511 & 0.381 & 0.39 & 1.02\\
      & [3.16, 3.42] & [0.334, 0.361] & [0.499, 0.528] & [0.366, 0.401] & [0.35, 0.46] & [0.90, 1.12]\\ \hline
     Cat & 3.24 & 0.366& 0.534& 0.407& 0.48& 0.83\\
      & [3.06, 3.42] & [0.352, 0.381] & [0.519, 0.551] & [0.388, 0.428]& [0.41, 0.58]& [0.68, 0.95]\\ \hline
     Treeshrews & 3.08 & 0.364 & 0.521 & 0.404 & 0.36 & 1.13\\
      & [2.99      3.16] & [0.359      0.370] & [0.514     0.528] & [0.396     0.411] & [0.34      0.39] & [1.05      1.19]\\ \hline
     Galago & 3.12& 0.363& 0.536& 0.396& 0.45& 0.85\\ 
      & [2.93, 3.27]& [0.345, 0.381]& [0.522, 0.556]& [0.375, 0.417]& [0.42, 0.52]& [0.71, 0.99]\\ \hline

    Ensemble Average & 3.14& 0.359 & 0.525 & 0.396 & 0.40 & 1.05  \\ \hline
    Common Design--CR& [3.09      3.19] & [0.344     0.357] & [0.506     0.522] & [0.387     0.399] & [0.37     0.42] & [0.99      1.11]   \\ \hline
    One Species--CR & [2.93, 3.42] & [0.334, 0.381]& [0.499, 0.556] & [0.366, 0.428] & [0.34, 0.58]& [0.68, 1.19]\\ \hline
    \end{tabular}\\
    \end{small}
    \end{adjustwidth}
    \vspace{4pt}
    \caption{The six orientation domain layout parameters characterizing the common design. Values were calculated with the code provided in the supplemental material and intervals indicate $95\%$ bootstrap confidence intervals. Also shown is the grand average and the associated one species and common design consistency ranges (CR).\label{commdsgn}}
\end{table}
%
%
%
\subsection*{The statistical wiring model}
\begin{figure}
\begin{center}
\includegraphics[width=\linewidth]{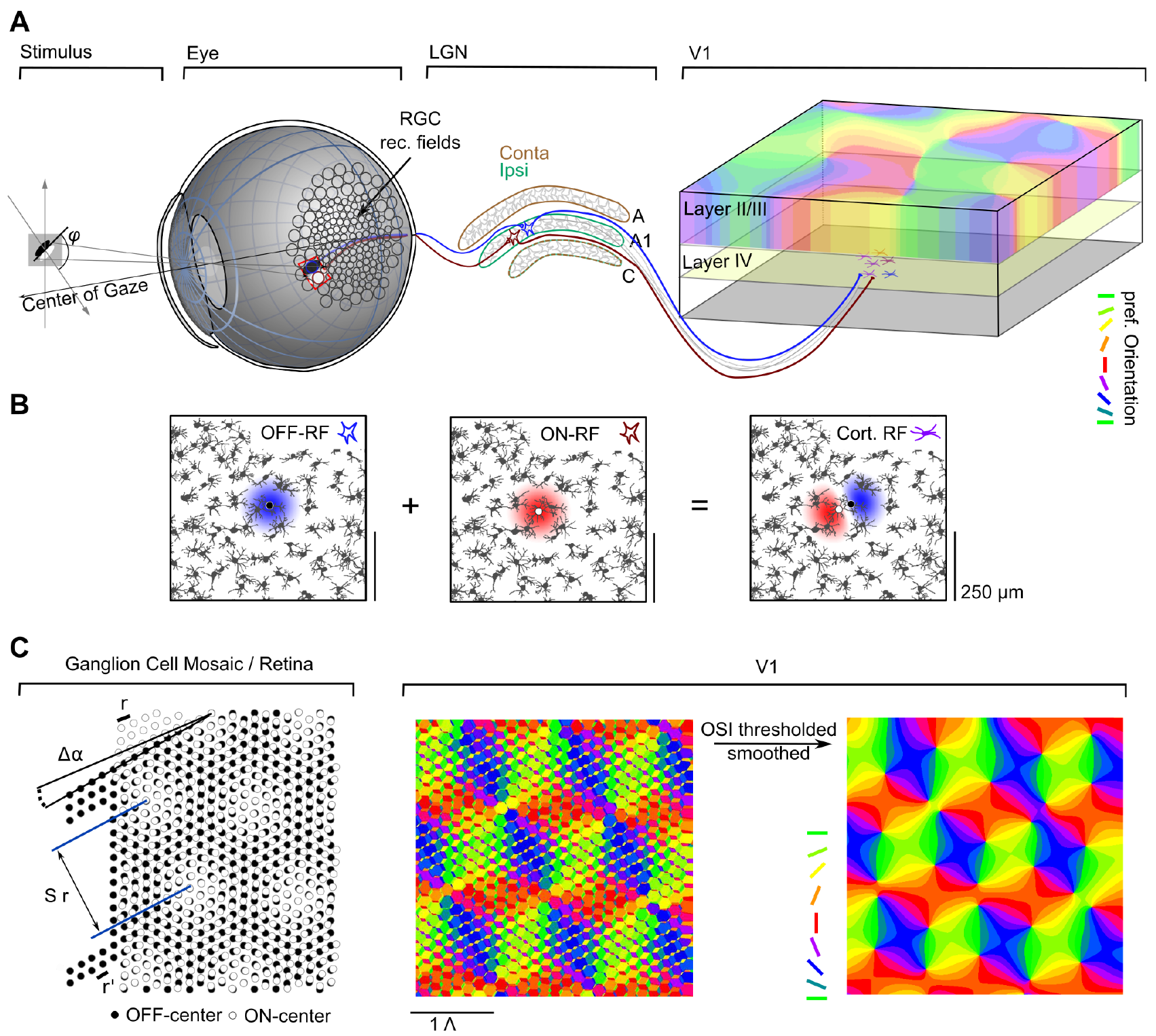}
\end{center}
\caption{\textbf{Early visual pathway, RGC dipoles, and Moir\'{e} interference of RGC mosaics} \textbf{A} Schematic illustration of the early visual pathway following the organization in the cat (see text for details). \textbf{B} Orientation selective receptive fields arise through summation of two adjacent rotationally symmetric retinal/LGN receptive fields (RGC dipole). Shown are ON and OFF center mosaics from cat retina \cite{Wassle1981}. Colors indicate ON (red) and OFF (blue) regions of a receptive field in the LGN (left and middle) and V1 (right). For illustration, the RGC mosaic is overlaid and the two RGCs whose RFs are summed are shown as black and white dots.
\textbf{C} Left: Moir\'{e} interference between a hexagonal ON (white dots) and OFF (black dots) RGC lattice with relative orientation $\Delta \alpha$ and lattice constants $r$ and $r'$ (black bars) creates a Moir\'{e} pattern with lattice constant $S\cdot r$. Middle: sampling from this RGC mosaic as described in B (and text) yields a periodic orientation preference pattern through Moir\'{e} interference. Right: Model layout predicted by the statistical wiring model, obtained by thresholding and smoothing the pattern in the middle (see text).}
\label{fig_1}
\end{figure}
The statistical wiring model formalizes the idea that the spatial progression of orientation preference domains arises from the spatial distribution of RGC receptive fields on the retina via feedforward wiring. \textbf{Fig.~\ref{fig_1}A} shows a simplified schematics of the early visual pathway in the cat \cite{Alonso2001, Usrey1999, Ferster1996, Reid1995, Thompson1994, Shou1989, Levick1980, Levay1976, Hammond1974, Hubel1962, Boycott1974, Hubel1963, Hubel1961, Hubel1959, Ohki2006}, from the retina to layer IV of V1. A stimulus is focussed onto the retina through the cornea and lens, is sampled by RGC RFs and transmitted to the LGN. LGN neurons project to stellate cells in layer IV of V1, whose responses are orientation tuned. Orientation tuning varies smoothly across the cortical surface.  \\
In the model, RGCs are assumed to be mono-synaptically connected one-to-one to relay cells in the LGN. Thus, the receptive fields of LGN neurons are similar to those of RGCs and the spatial arrangement of ON/OFF receptive fields of relay cells in the LGN mirrors the RGC receptive field mosaic. Neurons in the model visual cortex linearly sum inputs of LGN neurons (\textbf{Fig.~\ref{fig_1}B}). Their spatial receptive fields and orientation preferences are assumed to solely depend on the spatial arrangement of their afferent inputs. V1 neurons are assumed to receive dominant inputs from a small number of geniculo-cortical axons. Most of them sample from a single pair of ON/OFF RGCs, a so-called RGC dipole (\textbf{Fig.~\ref{fig_1}B}). The neuron's receptive field then consists of one ON and one OFF subregion and its response to edge-like stimuli is tuned to an edge orientation \textit{orthogonal} to the RGC-dipole vector (\textbf{Fig.~\ref{fig_1}B}). Within a mosaic of ON and OFF center RGCs, many such dipoles 
are present and the spatial arrangement of dipoles on the retina determines how tuning properties, e.g. the preferred orientation, change along a two-dimensional sheet parallel to the layers of the visual cortex. If ON and OFF RGCs are positioned on hexagonal lattices, the model predicts that a hexagonal pattern of orientation preference can arise through Moir\'e interference (MI) between the two lattices (\textbf{Fig.~\ref{fig_1}C}). \newline
Following \cite{Ringach2004, Paik2011}, we model RGC receptive fields using a Gaussian function $\text{GRF}_j(\mathbf{x})$ of width $\sigma_r$, localized at the center position $\mathbf{x}_j$:
\begin{eqnarray}
 \text{GRF}_j(\mathbf{x})=\pm\exp\left(-\frac{(\mathbf{x}_j-\mathbf{x})^2}{2\sigma_r^2}\right)\,,\label{rgcrf}
\end{eqnarray}
where $\mathbf{x}$ indicates position in retinal space. All subsequent results remain qualitatively unchanged if a biologically more realistic difference-of-Gaussians (see \cite{Enroth-Cugell1966}) is used. A plus or minus sign in Eq.~\eqref{rgcrf} indicates an ON or OFF center cell, respectively.
The receptive field $\text{RF}_\mathbf{y}$ of a visual cortical neuron at position $\mathbf{y}$ in the two-dimensional cortical sheet is obtained by summing several ganglion cell receptive fields with positive synaptic weights $w_j$:
\begin{eqnarray}
 \text{RF}_\mathbf{y}(\mathbf{x})=\sum_j w_j(\mathbf{y})\text{GRF}_j(\mathbf{x})\,.
 \label{summation}
\end{eqnarray}
The synaptic weights are chosen as
\begin{eqnarray}
 w_j(\mathbf{y})=\exp\left(-\frac{(\mathbf{x}_j-\mathbf{y})^2}{2\sigma_s^2}\right)\,.
 \label{weight_sum}
\end{eqnarray}
The parameter $\sigma_s$ sets the range from which a V1 neuron receives retinal inputs, $\mathbf{x}_j$ denotes the center of an RGC receptive field. According to Eq.~\eqref{summation} the spatial distribution of RGC locations determines how response properties change across cortex. For $\sigma_s$ smaller than the lattice spacing, each cortical cell receives substantial input only from a very small number of ganglion cells. Inputs received by most cortical cells are dominated by one ON and one OFF center RGCs (see inset in \textbf{Fig.~\ref{fig_1}A}), forming an RGC dipole. The small $\sigma_s$ regime is thus generally referred to as the dipole approximation of the model. While the dipole approximation leads to the robust emergence of simple-cell receptive fields with one (ON, OFF) or two (ON-OFF) subfields in the model V1 layer, it is worth mentioning that simple cells in cat and macaque monkey sometimes have more than two aligned, regularly spaced subfields (e.g. ON-OFF-ON or OFF-ON-OFF) (see \cite{
Jones1987a, Ringach2002a}). In the dipole approximation of the statistical wiring model, such simple-cell RFs almost never occur. While the model as defined above implements a deterministic wiring scheme, it represents a simplification of a more detailed formulation of the statistical connectivity model proposed in \cite{Ringach2004}. In the more detailed formulation, the synaptic weights between the cortical units and the retina/LGN are chosen at random from a Gaussian distribution with the shape given in Eq.~\eqref{weight_sum}. Ringach established in \cite{Ringach2004} that the spatial structure of the resulting domain layouts for the detailed and simplified model are nearly identical. We therefore refer to the model as statistical connectivity model.\\
We used the linear response assumption \cite{Heeger1991,Ferster1994} to determine cortical stimulus responses. A response $R$ of a cortical neuron is modeled by the inner product between its receptive field $\text{RF}_\mathbf{y}(\mathbf{x})$ and the stimulus, in our case an illumination pattern $\text{L}(\mathbf{x})$:
\begin{equation}
R_\mathbf{y}=\int \mathrm{d}^2 \mathbf{x}\, \text{RF}_\mathbf{y}(\mathbf{x})\,\text{L}(\mathbf{x})\,.
\label{response}
\end{equation}
Because $R_\mathbf{y}$ can become negative, a firing rate $f$ of the cortical neuron is then defined through a static nonlinearity, e.g. half-wave rectification \cite{Heeger1991}. For the purposes of the present study, this nonlinearity can be neglected assuming that it does not alter core properties of the receptive field such as orientation preference and spatial frequency preference \cite{Ringach2004, Ringach2007}. \\
We derived close-form expressions for the pattern of cortical receptive fields across V1 
that arises through Moir\'e interference in the case that ON and OFF center cells are localized on perfectly hexagonal lattices with different lattice constants $r$ and $r'$ and relative angle $\alpha$ between the lattices (see \textbf{Fig. \ref{fig_1}C}). Detailed derivations are provided in Methods, along with closed-form expressions for receptive fields, the frequency response of orientation selective neurons, and their spatial organization.\\
\textbf{Fig.~\ref{fig_3}A} depicts the analytically calculated orientation preference pattern generated through Moir\'e interference between two hexagonal ON/OFF RGC mosaics. Iso-orientation domains are organized in fine-grained parcellations on small scales and repeat in a hexagonal pattern on a larger scale $\Lambda$. The larger scale is the predicted column spacing of the orientation domain layout (see Methods), and model parameters are chosen such that $\Lambda\approx 1\text{mm}$ as experimentally measured (see Methods and \cite{Paik2011, Paik2012} for details). The scale of the small parcels therefore is $<200\text{\textmu m}$. A magnified view of a small region of the domain layout is provided in \textbf{Fig.~\ref{fig_3}B} along with three analytically determined cortical receptive fields at closely spaced locations roughly 100 \textmu m apart from each other. These receptive fields highlight that individual parcels contain highly tuned units with vastly different preferred orientations. This means 
that orientation preference changes abruptly on scales $<200\text{\textmu m}$ in the predicted patterns. Clearly, these features distinguish the obtained pattern of orientation preferences from the experimentally observed domain layouts. While orientation selectivity in V1 exhibits some small scale scatter within orientation domains \cite{Hetherington1999, Nauhaus2008}, two-photon imaging suggests that orientation preferences progresses rather smoothly across the cortical surface \cite{Ohki2005, Ohki2006}.
\subsection*{Orientation Preference Maps from crystalline RGC mosaics}
%
%
\begin{figure}
\begin{center}
\includegraphics[width=\linewidth]{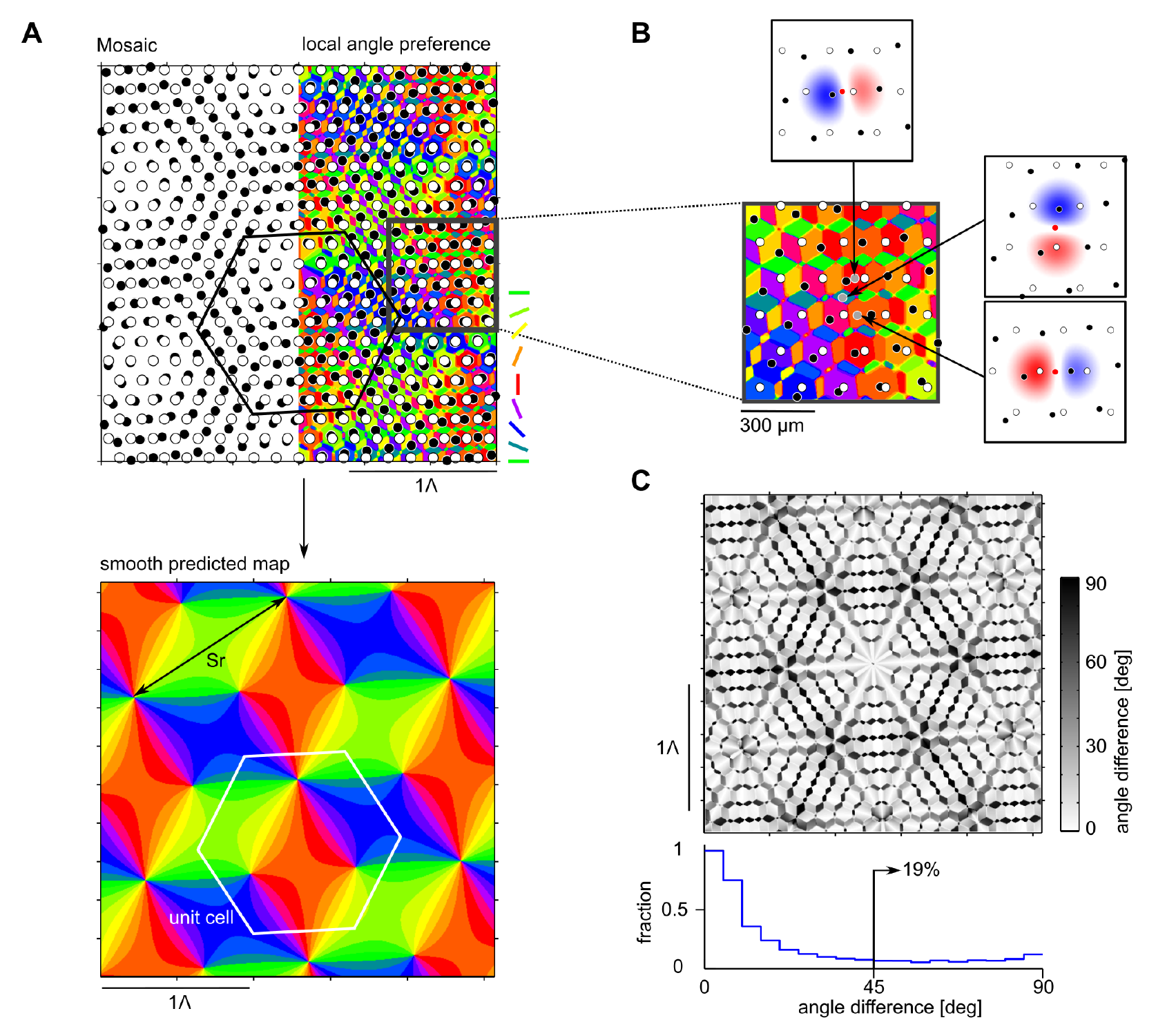}
\end{center}
\caption{\textbf{Receptive fields and iso-orientation domains in the Moir\'e interference model}. \textbf{A} Top: Moir\'e interference between two RGC mosaic (left) with ON and OFF center RGCs illustrated as white and black dots. The corresponding orientation domain layout (Eq.~\eqref{fullopm} in Methods) is shown on the right with the mosaic overlaid. Bottom: low frequency contribution of the domain layout. Black arrows indicate the lattice constant $S\cdot r$ of the Moir\'{e} pattern, white hexagon indicates the unit cell of the domain layout.
\textbf{B} Inset of the layout shown in A with RFs of three closely spaced neurons. Scale bar indicates distance on the retina.
\textbf{C} Circular distance (see text) between the preferred angles of unfiltered and low-pass filtered domain layouts shown in A top and bottom. Bottom: Histogram of differences in preferred angles. Model parameters: $\sigma_r=70\text{ \textmu m}$, $\sigma_s=20\text{ \textmu m}$, lattice constants $r=r'=170\text{ \textmu m}$, and a relative angle $\Delta\alpha=7^\circ$ leading to a scaling 
factor of $S=8.2$ (Eq. \eqref{scaling_factor}), as proposed in \cite{Paik2011,Paik2012}.}
\label{fig_3}
\end{figure}
\begin{figure}
\begin{center}
\includegraphics[width=\linewidth]{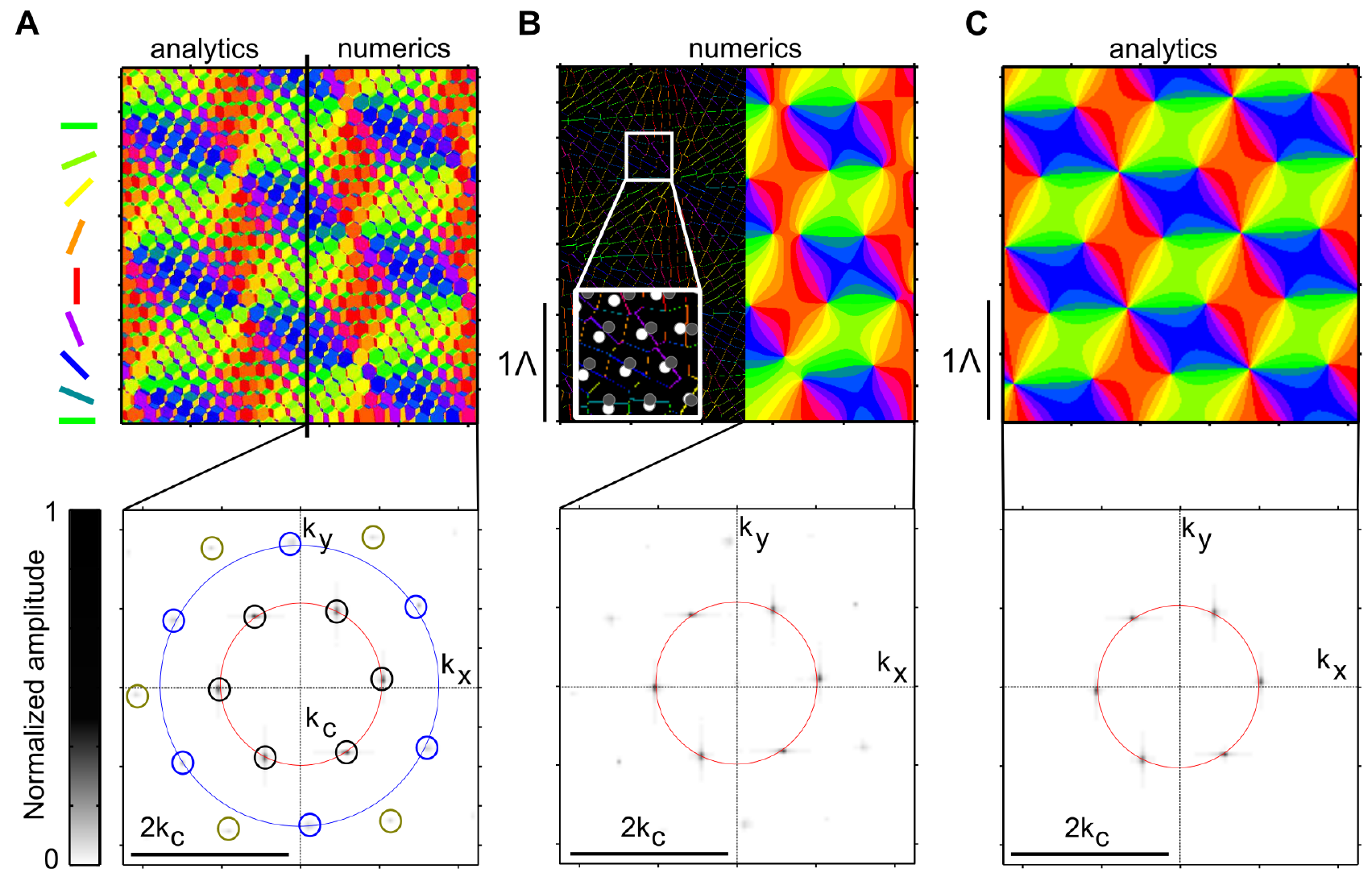}
\end{center}
\caption{\textbf{Comparison of analytically and numerically obtained solutions of the Moir\'e interference model.} \textbf{A} Top: unfiltered Moir\'e interference patterns. Black line separates analytical (left, see Eq.~\eqref{fullopm}) from numerical result (right). Bottom: amplitude spectrum of numerically obtained Moir\'e interference patterns. Red circle marks $k_c$ (cf. Eq.~\eqref{kcmap}), blue circle indicates $\sqrt{3}k_c$. Note the high frequency contributions, indicated by the small yellow circles. \textbf{B} Top: Numerically obtained Moir\'e interference pattern after thresholding for cells with $\text{OSI}>0.25$ (left, see Methods for the OSI definition used) and subsequent smoothing (right, see text). Inset shows a magnified region of the OSI-filtered domain layout together with the RGC mosaic from which the neurons sample. Bottom: amplitude spectrum of the numerically obtained thresholded and smoothed layout. Red circle indicates $k_c$ (cf. Eq.~\eqref{kcmap}). 
\textbf{C} Orientation domain layout (top) and amplitude spectrum (bottom) obtained by calculating the lowest spatial frequency contributions of the layout in A (Eq.~\eqref{moiremodes}). All model parameters as in \textbf{Fig.~\ref{fig_3}}.}
\label{fig_4}
\end{figure}

Paik \& Ringach implicitly assumed that random feedforward wiring from the retina/LGN to V1 effectively results in a \textit{smoothed} version of the dipole layout	 (see \textbf{Fig.~\ref{fig_1}C}). To extract this smooth pattern of orientation preferences from the statistical connectivity model, they adopted a two-step procedure to suppress the small-scale variation in the Moir\'{e} interference pattern: First, locations with orientation selectivity index (OSI) larger than a threshold value are determined \cite{Ringach2004,Ringach2007, Paik2011, Paik2012}. Second, the orientation selectivity of all other location is set to zero. The resulting layout is then filtered with a Gaussian lowpass filter resulting in continuous and smooth array of iso-orientation domains \cite{Ringach2007, Paik2011, Paik2012}. \\
We find that the thresholding/smoothing procedure effectively extracts the dominant lowest spatial frequency Fourier components of the Moir\'{e} interference pattern. As derived in Methods, the lowest spatial frequency contribution to the Moir\'{e} interference pattern consists of six Fourier modes with identical amplitude and wave number
\begin{eqnarray}
 k_c&=&\frac{4\pi}{\sqrt{3}rr'}\sqrt{r^2+r'^2-2rr'\cos(\Delta\alpha)},\label{kcmap}
\end{eqnarray} 
Here, $r,r'$ denote to the lattice constants and $\Delta\alpha$ the angle between the hexagonal ON/OFF lattices. (\textbf{Fig.~\ref{fig_1}C}). The smooth orientation domain layout resulting from Moir\'e interference can therefore be summarized in a complex-valued field $z(\mathbf{y})$ composed of six planar waves with wave numbers $\mathbf{k}_j$ and fixed phase factors $u_j$,
\begin{eqnarray}
  z(\mathbf{\mathbf{y}})=\sum_{j=1}^6\exp(\mathfrak{i}\mathbf{k}_j \mathbf{\mathbf{y}})\cdot u_j\,.\label{compfield}
\end{eqnarray}
The pattern of preferred orientations across the cortical coordinate $\mathbf{y}$ is given by the phase of this complex-valued field as,
\begin{eqnarray}
 \vartheta_\text{pref}(\mathbf{y})=\frac{1}{2}\text{arg}\left(z(\mathbf{y})\right)\,.
 \label{moiremodes}
\end{eqnarray}
\textbf{Fig.~\ref{fig_3}A} (bottom) depicts $\vartheta_\text{pref}(\mathbf{y})$ as analytically determined. The pattern of pinwheels and iso-orientation domains is organized into a smooth hexagonal crystalline array. Interestingly, an identical layout of iso-orientation domains was constructed by Braitenberg et al. \cite{Braitenberg1979} based on an the idea that orientation preference is generated by discrete centers of inhibition in V1. It was also found by Reich et al. to solve a symmetry defined class of models for the self-organization of iso-orientation domains \cite{Reichl2012a,Reichl2012b}.  \textbf{Fig.~\ref{fig_3}C} shows the differences in preferred angle\footnote{With $\Delta(\mathbf{x})=\vartheta_1(\mathbf{x})-\vartheta_\text{pref}(\mathbf{x})$, the difference $d(\mathbf{x})$ between the two preferred angles is $d(\mathbf{x})=\frac{1}{2}\textnormal{abs}\left(\textnormal{arg}\left(e^{2 \mathfrak{i} \Delta(\mathbf{x})}\right)\right)$} between the unfiltered domain layout of the Moir\'e 
interference pattern and its low frequency contribution, together with a histogram of the differences.  The bimodal shape of the histogram indicates that the orientation preference of a large fraction of cortical locations differs substantially between unfiltered and smoothed layout. Roughly one fifth of all locations exhibit differences of orientation preferences of more than $45^\circ$. \newline
To compare our mathematical expression for the column spacing of the orientation domain layout to previous results, Eq.~\eqref{kcmap} can be rewritten by introducing a parameter $\beta$ representing the detuning between the two lattice constants in units of the lattice constant $r' \to (1+\beta) r$.
The expression for the column spacing becomes
\begin{eqnarray}
\Lambda_c\equiv\frac{2\pi }{k_c}=\frac{\sqrt{3}}{2}\cdot S\cdot r\,,
\label{sc}
\end{eqnarray}
where $S$ is the distance between two vertices of the Moir\'{e} pattern in units of $r$, called the scaling factor \cite{Blair2007, Amidror2009, Nishijima1964}
\begin{eqnarray}
 S=\frac{1+\beta }{\sqrt{\beta ^2+2(1-\cos(\Delta \alpha))(1+\beta )}}\,.
 \label{scaling_factor}
\end{eqnarray}
The difference between $S\cdot r$ and $\Lambda_c$ is displayed in \textbf{Fig.~\ref{fig_3}A} (bottom). Eqs.~\eqref{sc} and \eqref {scaling_factor} are identical to previous results for the spacing of hexagonal Moir\'{e} patterns derived via geometrical considerations \cite{Amidror2009, Nishijima1964}. \newline
Using these explicit expressions for the iso-orientation domain layout and its column spacing $\Lambda_c$, we first evaluated the central quantity of the common design - the pinwheel density, i.e. the number of pinwheels per unit area $\Lambda_c^2$. Within each unit cell of area $A=\frac{\sqrt{3}}{2}\left(S\cdot r\right)^2$, there is one ``double pinwheel'' of topological charge $1$, around which each orientation is represented twice, and two pinwheels of topological charge $\pm\frac{1}{2}$. With $\Lambda_c^2=\frac{3}{4}(S\cdot r)^2$ and counting the pinwheel with charge $1$ as two pinwheels (see below), the pinwheel density is
\begin{eqnarray}
\rho=\left(2+2\cdot1\right)\cdot\frac{\Lambda_c^2}{A}=2\sqrt{3}\approx3.46.
\label{pw_dens_analytics}
\end{eqnarray}
Notably, this value is outside of both, the common design consistency range and the single species consistency range for the experimentally measured pinwheel densities (cf. Table 1). Since the statistical connectivity model for perfectly hexagonal RGC mosaics results in a periodic array of pinwheels, all three nearest neighbor distance distributions of pinwheels are sharply peaked (see also Supplementary Material of \cite{Kaschube2010}) and, thus, in disagreement with the distributions experimentally observed (cf. \textbf{Fig.~\ref{fig_0}}).\newline
We compared these analytical results to numerically evaluated Moir\'{e} interference patterns (\textbf{Fig.~\ref{fig_4}}). The fine-grained layouts of numerically and analytically obtained unfiltered layouts are almost indistinguishable (cross-correlation coeff. 0.9, \textbf{Fig.~\ref{fig_4}A} top). This confirms the analytical treatment and indicates accuracy of the numerical implementation. A hierarchy of discrete spatial frequency contributions is apparent in amplitude spectra of both domain layouts (\textbf{Fig.~\ref{fig_4}A} (bottom)).  The peaks at larger spatial frequencies in \textbf{Fig.~\ref{fig_4}A,B} are localized at $\sqrt{3}k_c$ as analytically predicted (see Methods). \\
To numerically generate the smoothed array of orientation domains, the layout in \textbf{Fig.~\ref{fig_4}A} (top) was thresholded ($\text{OSI}>0.25$, see Methods) and subsequently smoothed with a Gaussian lowpass filter (\textbf{Fig.~\ref{fig_4}B} top) \cite{Ringach2007, Paik2011}.  In general, strongly tuned locations are those exactly between ON-OFF RGC pairs (\textbf{Fig.~\ref{fig_4}B}, inset). \textbf{Fig.~\ref{fig_4}C} depicts the crystalline pinwheel arrangement of the analytically calculated smoothed layout (Eq.~\eqref{moiremodes}) as well as the six dominant low frequency Moir\'{e} modes in the amplitude spectrum. While the numerically obtained layouts and its analytical approximation are similar (cross-correlation coeff. 0.6), one major difference can be observed: the pinwheel of topological charge $1$ is replaced by two pinwheels of topological charge $\frac{1}{2}$ in the numerically obtained layouts, along with subtle deformations of adjacent orientation domains (compare \textbf{Fig.~\ref{fig_4}B 
and C}). To see why this is the case, we note that the pinwheel with charge $1$ in the analytically calculated pattern (Eq.~\eqref{moiremodes}) arises from a zero of the field $z(\mathbf{x})$ (Eq.~\eqref{compfield}) with multiplicity two. A phase singularity of a complex-valued field arising from a zero with multiplicity $N>1$ is structurally unstable and unfolds upon generic infinitesimal perturbations into N closely spaced singularities of multiplicity one \cite{Berry2001}. The numerical procedure of discretizing V1 unit positions on a numerical grid, OSI thresholding, and smoothing realizes such a perturbation and this explains why in the numerical solutions the pinwheel of charge $1$ unfolds into two adjacent pinwheels of charge $\frac{1}{2}$.
%
%
%
%
%
%
%
%
%
\subsection*{The impact of spatially uncorrelated disorder in RGC position}
\begin{figure}
\begin{center}
\includegraphics[width=\linewidth]{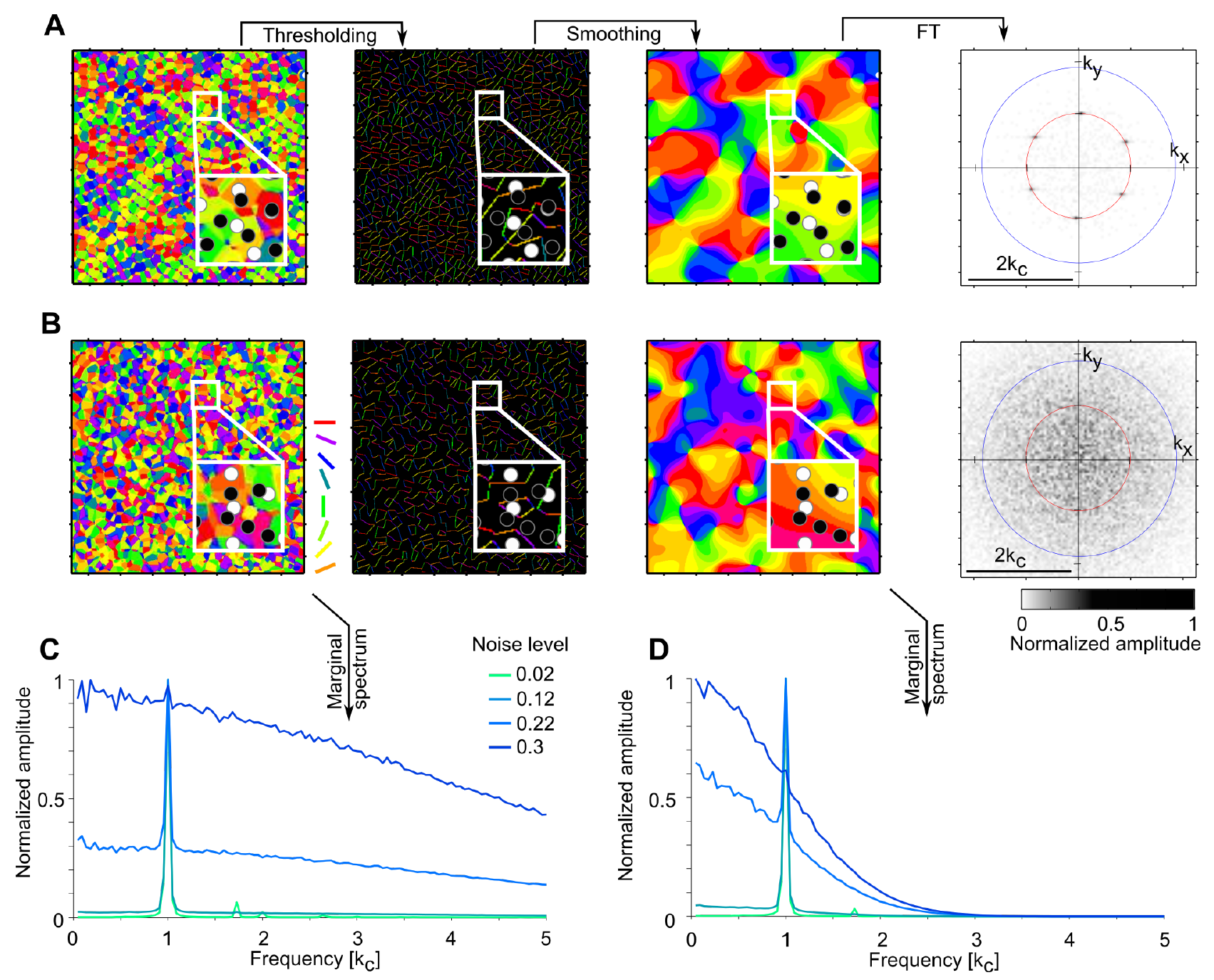}
\end{center}
\caption{\textbf{Spatially uncorrelated position disorder in hexagonal RGC mosaics induce broadband noise in iso-orientation domain layouts.} \textbf{A} Numerically calculated orientation domain layouts with disorder strength $\eta=0.12$. From left to right: Moir\'e interference pattern, filtered Moir\'e interference pattern ($\text{OSI}>0.25$), smooth layout and the smoothed layout's amplitude spectrum. Insets show magnified regions. Circles in the amplitude spectrum mark $k_c$ (red) and $\sqrt{3}k_c$ (blue) (cf. Eq. \eqref{kcmap}). 
\textbf{B} As A but for a higher disorder strength $\eta=0.30$. 
\textbf{C} Radially averaged normalized amplitude spectra of the orientation domain layouts for different disorder strengths. The fluctuation strength is color coded (legend). x-axis is given in units of $k_c$ (cf. Eq. \eqref{kcmap}).
\textbf{D} As C but for the smoothed layouts. All other model parameters as in \textbf{Fig.~\ref{fig_3}}.}
\label{fig_5}
\end{figure}
\begin{figure}
\centering
\includegraphics[width=\linewidth]{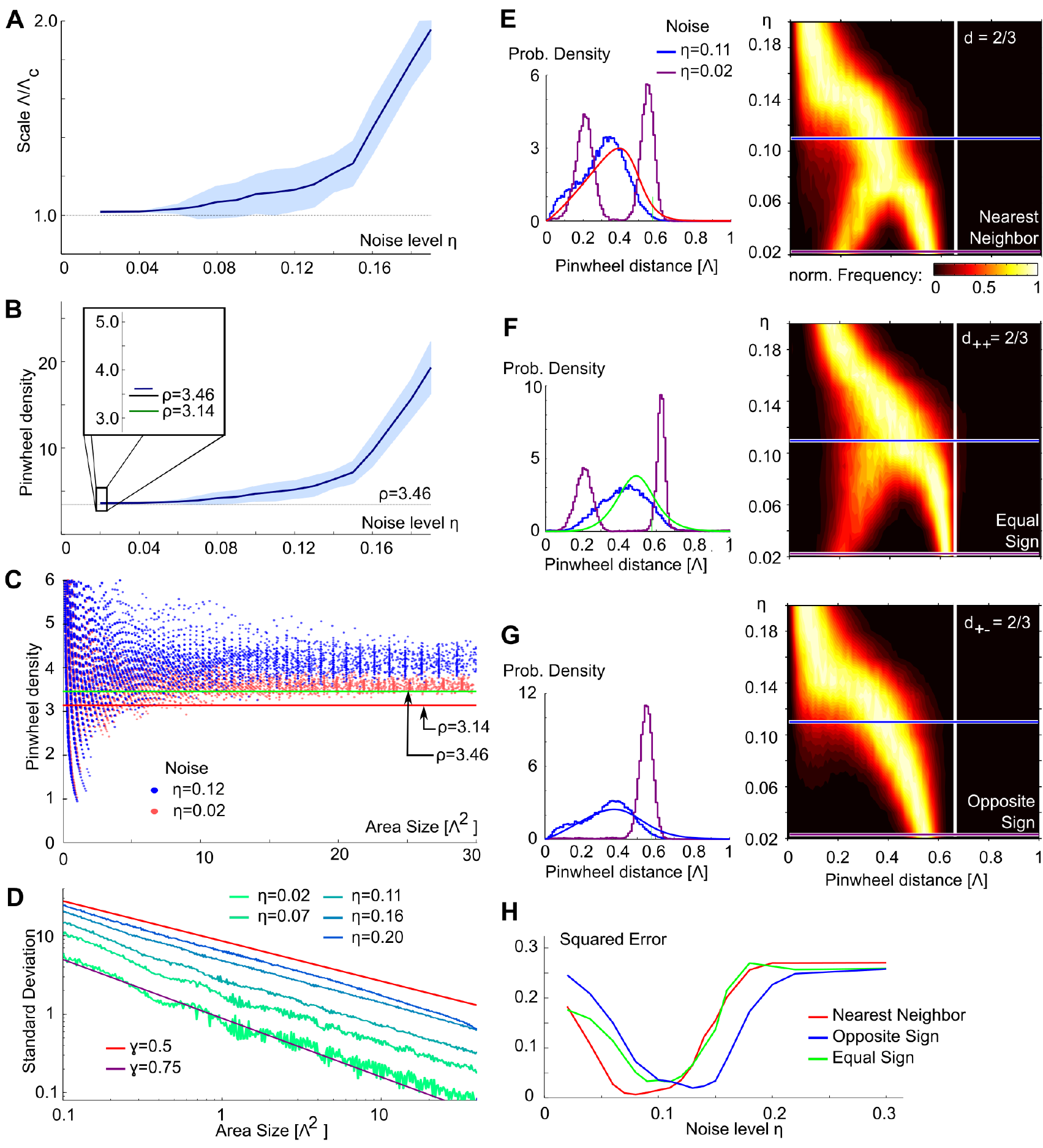}
\caption{\label{fig_7}\textbf{Pinwheel statistics in the Moir\'e interference model}. 
\textbf{A} Column spacing $\Lambda$ estimated by the wavelet method compared to the Moir\'{e} scale $\Lambda_c$ for different disorder strength. 
\textbf{B} The pinwheel density $\rho$ as function of disorder strength. Dashed line shows theoretically predicted value $\rho=2\sqrt{3}$. Dark green line in inset shows experimentally observed mean value $\rho=3.14$. 
\textbf{C} Pinwheel density in circular regions of increasing area for $\eta=0.12$ (blue) and $\eta=0.02$ (red). Lines show theoretically predicted and experimentally observed values as in B inset. 
\textbf{D} The standard deviation of pinwheel density estimates for increasing subregion size. Red line shows a power law fit to the experimental data ($\gamma = 0.5$,\cite{Kaschube2010}), purple line indicates a fit to the perfectly ordered hexagonal pinwheel arrangement ($\gamma = 0.75$,\cite{Kaschube2010, Reichl2012a,Reichl2012b}).
\textbf{E} Nearest neighbor (NN) distances for pinwheels irrespective of topological charge. Left: distributions for two disorder strengths ($\eta = 0.11$, blue; $\eta = 0.02$, purple) and the experimental data (red). Right: Distributions for different disorder strengths. Color encodes the (normalized) fraction of pinwheels at this distance. Blue and purple lines indicate disorder strengths shown on the left. The white line marks the theoretically predicted distance of NN pinwheels ($2/3\Lambda_c$) for vanishing disorder \cite{Reichl2012a,Reichl2012b}.
\textbf{F} same as E for pinwheels of equal charge, data green curve. 
\textbf{G} same as E for pinwheels of opposite charge, data blue curve. 
\textbf{H} squared deviation of NN distance distributions to the experimental estimates (shown in E-G, left) as function of disorder strength. All other model parameters as in \textbf{Fig.~\ref{fig_3}}.}
\end{figure}
\begin{figure}
\centering
\includegraphics[width=\linewidth]{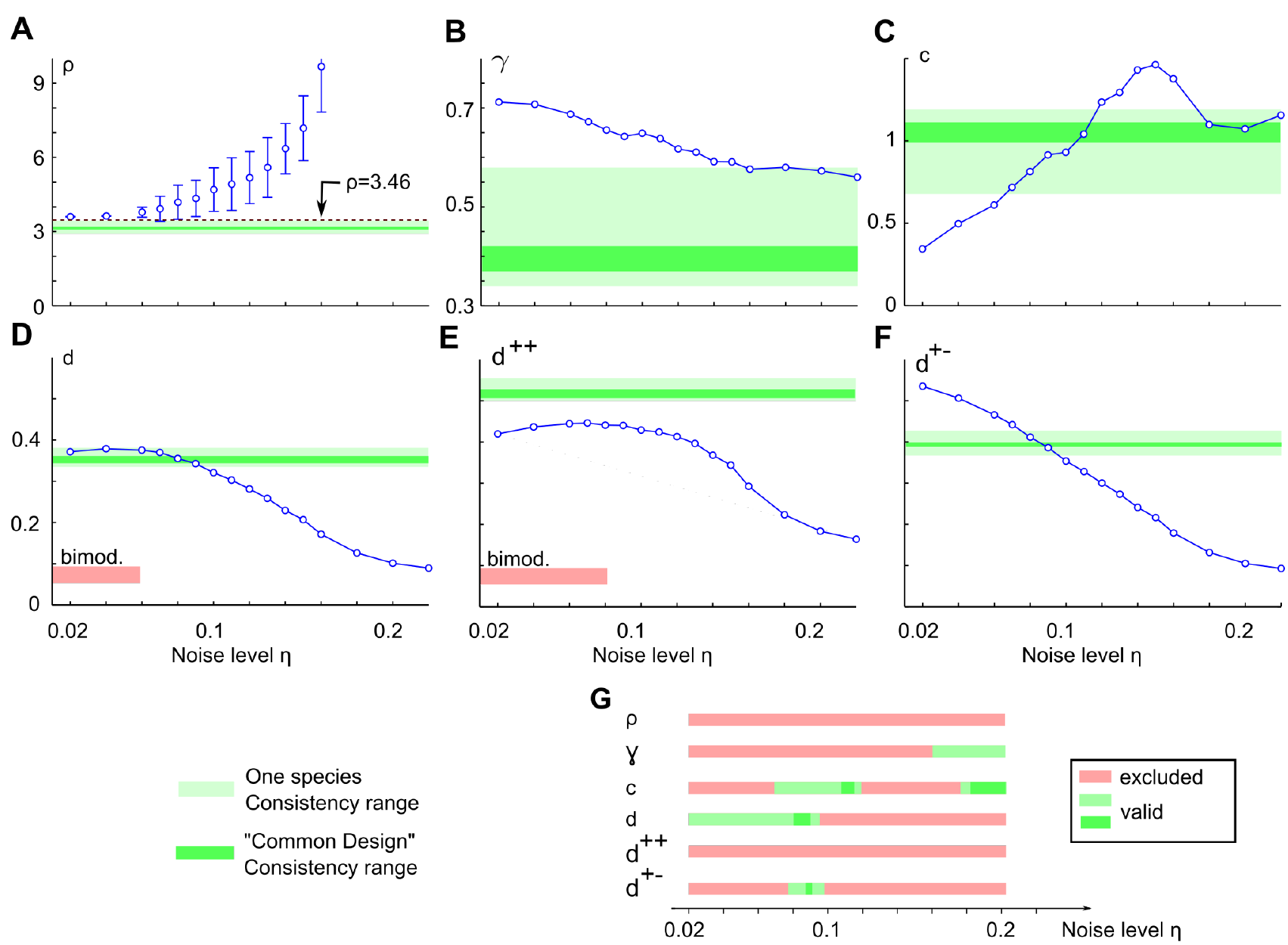}
\caption{\label{fig_8}\textbf{Pinwheel statistics of the disordered Moir\'e interference model fail to match V1 functional architecture.} \textbf{A} Pinwheel density as a function of disorder strength in the statistical connectivity model with noisy hexagonal mosaics. Error bars indicate standard deviation around the mean for 20 model realizations, circles the mean. Green shaded areas indicate the range consistent with the experiments (see \textbf{Fig.~\ref{fig_0}}). 
\textbf{B} The variability exponent as function of disorder strength in comparison to the common design. \textbf{C} As B for the variability constant. \textbf{D} Mean nearest neighbor distance for pinwheels independent of topological charge in comparison to common design. \textbf{E} As D but for pinwheels of equal charge. \textbf{F} As D for pinwheels of opposite charge. \textbf{G} Summary of ranges of disorder parameters consistent with the common design. Disorder strengths larger than $0.3$ can be excluded by the lack of typical column spacing in the domain layouts (cf. \textbf{Fig.~\ref{fig_5}}). Note that there is no disorder strengths for which all features of the common design are reproduced.}
\end{figure}

So far, we have studied the idealized situation of iso-orientation domains induced by perfectly ordered hexagonal RGC mosaics. RGC mosaics in the eye, however, are not perfectly hexagonal but exhibit substantial spatial irregularity \cite{Wassle1981, Field2007}. Therefore, we next turned to numerically investigate the statistical connectivity model with hexagonal RGC mosaics subject to Gaussian disorder in RGC position as previously described \cite{Paik2011,Paik2012}. The effect of ganglion cells displaced by Gaussian distributed offsets with standard deviation $\sigma=\eta\cdot r$ is illustrated in \textbf{Fig.~\ref{fig_5}}. The parameter $r$ is the lattice constant and $\eta$ is the disorder strength. \textbf{Fig.~\ref{fig_5}} shows the unfiltered orientation domain layout (far left), the layout thresholded for cells with an $\text{OSI}>0.25$ (left), the smoothed thresholded layout (right) as well as its amplitude spectrum (far right), numerically obtained for $\eta=0.
12$. As in the perfectly ordered case, the unfiltered layout of the noisy Moir\'e interference model 
exhibits a substantial scatter of orientation preferences across small scales. For a disorder strength of $\eta=0.12$ (\textbf{Fig.~\ref{fig_5}A}), the domain layout is still dominated by the six lowest spatial frequency Moir\'{e} modes also present in the perfectly ordered system. For a disorder strength of $\eta=0.3$ (\textbf{Fig.~\ref{fig_5}B}), the amplitude spectrum  (\textbf{Fig.~\ref{fig_5}B}, far right) lacks any indication of theses Moir\'{e} modes indicating that Moir\'{e} interference no longer takes place. As a consequence the resulting layouts of iso-orientation domains lack a typical column spacing. \\
To characterize the model orientation domain arrangements, we first calculate amplitude spectra for both, unfiltered and smoothed layouts (\textbf{Figs.~\ref{fig_5}A, B}),
\begin{eqnarray}
 |\mathcal{R(\mathbf{k})}| = \left|\int \mathrm{d}^2\mathbf{x}\, z(\mathbf{x}) e^{\mathfrak{i}\, \mathbf{k}\mathbf{x}}\right| \text{ where } z(\mathbf{x})=e^{2\mathfrak{i}\,\vartheta_\text{pref}(\mathbf{x})}.
\end{eqnarray}
Normalizing and radially averaging yields the so-called marginal amplitude spectrum (\textbf{Figs.~\ref{fig_5}C, D}),
\begin{eqnarray}
f(k)=\frac{\int_0^{2\pi} \mathrm{d}\vartheta\, |\mathcal{R}(k\cos(\vartheta),k\sin(\vartheta))|}{ \text{max}_k \,\int_0^{2\pi} \mathrm{d}\vartheta\, |\mathcal{R}(k\cos(\vartheta),k\sin(\vartheta))|}.
\end{eqnarray}
The sharp peak at $k_c$ corresponds to the dominant Moir\'{e} mode indicating that orientation domain layouts exhibit a typical column spacing. For increasing disorder, the relative levels of peak height to background decreases while the peak width remains small. As expected, marginal amplitude spectra of unfiltered and the smoothed layout mainly differ in the strength of background components. The flat amplitude spectrum of the unfiltered iso-orientation domain layouts for large disorder strength is transformed into a Gaussian amplitude spectrum by the lowpass filtering. Based on this assessment, the disorder strength $\eta$ has to be smaller than $0.3$ to ensure that layouts exhibit a typical spacing between adjacent iso-orientation domains.\\
We next systematically evaluated the core layout parameters of the common design -- pinwheel density and pinwheel nearest neighbor distance distributions for the statistical wiring model with disordered hexagonal RGC mosaics. To compare the model predictions with experiments, we estimated the column spacing of the model orientation domain layouts as well as pinwheel layout parameters using the exact same methods that we applied to the experimental data (see Methods). For weak disorder, column spacing estimates closely match the theoretical prediction $\Lambda_c$ (\textbf{Fig.~\ref{fig_7}A}), confirming the accuracy of the wavelet method. For disorder strengths larger then 0.12, Moir\'{e} modes are no longer the dominant spatial frequency contribution in the model layouts and the estimated column spacing increases with disorder strength.\\
Having estimated the column spacing, we analyzed model orientation domain layouts with respect to the common design parameters (\textbf{Fig.~\ref{fig_7}B-D}). As expected, pinwheel densities approach the analytical predicted value of $2 \sqrt{3}$ for weak disorder (\textbf{Fig.~\ref{fig_7}B}) and increase with increasing disorder strength. This increase is largely caused by the increase in the estimated column spacing (\textbf{Fig.~\ref{fig_7}A}) and does not involve a massive generation of additional pinwheels for larger disorder strength. We next calculated the standard deviation of pinwheel densities as a function of the area A of randomly selected subregions of the iso-orientation domain layouts. Generally, the standard deviation's decay with subregion size followed a power law with increasing area size, with larger exponents for weak disorder (\textbf{Fig.~\ref{fig_7}D}).\\
\textbf{Fig.~\ref{fig_7}E,F} show a complete characterization of pinwheel nearest neighbor (NN) distance distributions of the noisy Moir\'e interference model. Histograms for NN distances for arbitrary charge are bimodal for weak disorder (\textbf{Fig.~\ref{fig_7}E}). The peak at smaller NN distances results from the unfolding of pinwheels with topological charge $1$ into two adjacent pinwheels of topological charge $1/2$ for finite disorder strength (see above and \textbf{Fig.~\ref{fig_3},\ref{fig_4}}). For the same reason, the NN distance histogram for pinwheels of identical topological charge is also bimodal (\textbf{Fig.~\ref{fig_7}F}). With increasing disorder strength, both distributions become unimodal (\textbf{Fig.~\ref{fig_7}E,F} left). The NN distance distribution for pinwheels of opposite sign is unimodal for all parameter values, indicating that only very few additional pinwheel pairs are added to the pinwheels of the Moir\'{e} layout for small and intermediate disorder strengths (\textbf{Fig.~\ref{fig_7}G}). The overall decay in mean NN distance for strong disorder in all three histograms mostly reflects the increase in measured column spacing (see \textbf{Fig.~\ref{fig_7}A}). \\
Based on these results, we attempted to identify a disorder strength for which all NN pinwheel distance distributions resembled the experimental data. To this end, we calculated the squared error between the calculated histograms and the experimental data as a function of disorder strength (\textbf{Fig.~\ref{fig_7}H}). Smallest deviations from experimental data were obtained around $\eta\approx0.11$ for all three NN distance distributions.\\
%
%
%
\textbf{Fig.~\ref{fig_8}} summarizes all common design features determined for disordered Moir\'e interference model layouts as a function of disorder parameter and compares them to the experimentally observed values in tree shrew, galago, ferret, dark-reared ferrets, and cats. Light (dark) green shaded areas indicate the single species (common design) consistency ranges (see \textbf{Fig.~\ref{fig_0}}, cf. Table 1). 
With increasing disorder, pinwheel density of model layouts steadily increases from $\rho=2\sqrt{3}\approx 3.46$ (\textbf{Fig.~\ref{fig_8}A}) and always lies above the single species consistency range. Thus, the pinwheel density of model orientation domain layouts is inconsistent with pinwheel densities observed in all species. 
Next, we fitted the empirically observed power law, Eq.~\eqref{sdfun}, to the standard deviation of the pinwheel density estimate in increasing subregions of area $A$ (see \textbf{Fig.~\ref{fig_7}D}) \cite{Kaschube2010}. The variability exponent $\gamma$ is consistent with experiments for disorder levels exceeding $\eta=0.15$. The variability constant $c$ is monotonically increasing up to $\eta\approx0.15$ at which point the model domain layouts loose their typical column spacing (cf. \textbf{Fig.~\ref{fig_5}}) and the increasing pinwheel density $\rho$ causes a drop  (\textbf{Fig.~\ref{fig_8}C}). 
\textbf{Fig.~\ref{fig_8}D-F} displays the mean pinwheel NN distances as function of the disorder strength, all of which substantially decrease with increasing disorder strength. This can be attributed to the increasing mean column spacing of the domain layouts under increasing disorder (see \textbf{Fig.~\ref{fig_7}A}). Mean NN distances for weak disorder strength are close to the experimental data, but NN distance distributions for pinwheels of different topological charge and independent of topological charge are bimodal for weak disorder (\textbf{Fig.~\ref{fig_7}E,F}). The latter is clearly distinct from the experimental data (cf. \textbf{Fig.~\ref{fig_0}}, \cite{Kaschube2010}). \textbf{Fig.~\ref{fig_8}G} shows an overview of the consistency of model orientation domain layout parameters with the data for various disorder strengths. As can be seen, no strength of disorder results in layouts that are consistent with the common design for all layout parameters. Perhaps, even more surprising, pinwheel density 
and NN distance for pinwheels of the same sign are inconsistent with the individual values obtained for each species, no matter how the strength of disorder is chosen. 
%
%
%
%
\subsection*{Iso-orientation domain layouts from hexagonal RGC mosaics with spatially correlated disorder}
\begin{figure}
\centering
\includegraphics[width=\linewidth]{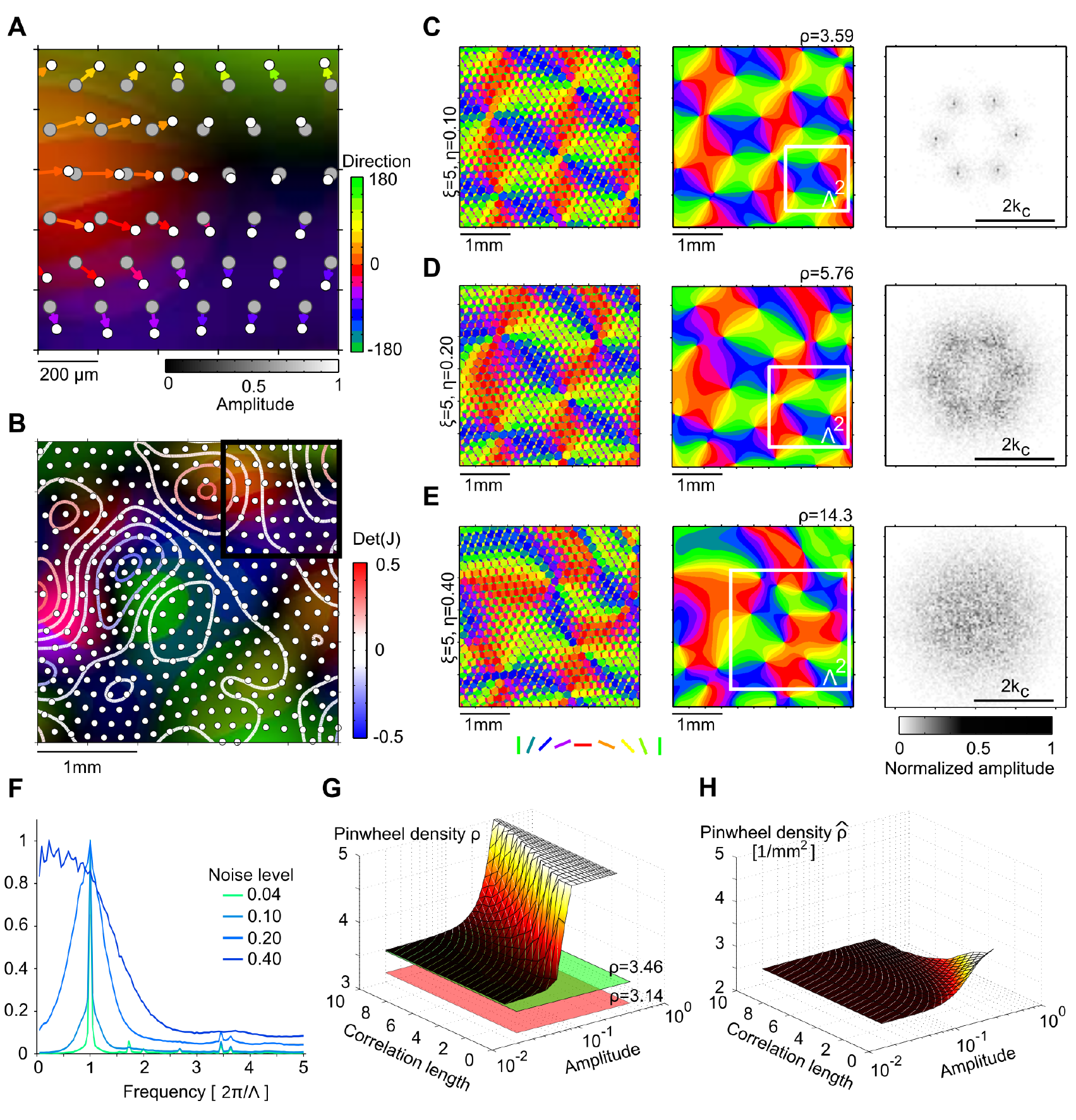}
\caption{\textbf{Impact of spatially correlated positional disorder in hexagonal RGC mosaics on iso-orientation domain layouts.}
\textbf{A} Generation of spatially correlated disorder in hexagonal RGC mosaics (see text and Methods). White dots mark ON RGC cells, gray dots marks the perfectly hexagonal ON-RGC lattice before distortion. Colors of the arrows and underlay indicate direction of RGC displacements (right colorbar). Magnitude of displacement is indicated by color saturation (bottom colorbar). Note that nearby RGCs move into similar directions and with similar magnitude. Scale bar indicates retinal distances, assuming ON/OFF lattice constant of $170\mu m$.
\textbf{B} Larger region of the mosaic shown in A (black square). Colors as in A. Contour lines mark lines of constant $\det(J(\mathbf{x}))$ (see text). Scale bar indicates retinal distances.
\textbf{C}-\textbf{E} Unfiltered model domain layout (left), thresholded and smoothed layout (middle) its amplitude  spectrum (right) for noise correlation length $\xi =5$ and different noise amplitudes ($\eta = 0.1$ (C), $\eta = 0.2$ (D), $\eta = 0.4$ (E)). White squares in middle panels indicate wavelength of pattern as measured by wavelet analysis (see Methods).  Scale bars indicate cortical distance with parameter choices as in \textbf{Fig.~\ref{fig_1}} and cortical magnification factor $\approx$1 (see \cite{Schottdorf2014}).
\textbf{F} Marginal amplitude spectra of smoothed domain layouts for $\xi = 5$ and different disorder strengths.
\textbf{G} Pinwheel density of model layouts as a function of disorder amplitude and correlation length. Red surface indicates experimentally determined value ($\rho = 3.14$), green surface indicates pinwheel density in the disorder-free case ($\rho = 2\sqrt{3}$). Values for $\rho \geq 5$ are drawn as a plane at $\rho = 5$.
\textbf{H} As G, but pinwheels per square millimeter $\hat{\rho}$. Note that the absolute number of pinwheels is largely constant. All other  model parameters as in \textbf{Fig.~\ref{fig_1}}.
\label{fig_9}}
\end{figure}

The above results show that the statistical connectivity model with hexagonal mosaics is unable to reproduce all features of the common design, even if spatially uncorrelated position disorder is imposed on the RGC positions. Whatever the source of disorder is that causes the irregularity in the RGCs' positions, it is plausible to assume that it is correlated on scales spanning several RGCs. Such spatial correlations would preserve the Moir\'{e} effect locally, yet generate spatial irregularity in orientation domain layouts.\\
To test whether correlated positional disorder can produce model arrangements of orientation domains that match experimental observations, we generalized the noisy hexagonal mosaics proposed in \cite{Ringach2004, Paik2011} to include spatial correlations. To obtain noisy hexagonal RGC mosaics with spatial correlations, we started with a hexagonal array of RGC positions. The position of each lattice point $\mathbf{x}_i$ was then shifted depending on its position according to $\mathbf{x}_i\to \mathbf{x}_i+\eta\,\mathbf{y}(\mathbf{x}_i)$. The shift $\eta\,\mathbf{y}(\mathbf{x})$ with amplitude $\eta$ for $\textbf{y}(\textbf{x})=(y_1(x),y_2(x))$ was chosen from a Gaussian random field with vanishing mean $\langle y_1(\mathbf{x})\rangle = \langle y_2(\mathbf{x})\rangle =0$, fixed standard deviation $\text{std}(y_1(\mathbf{x})) = \text{std}(y_2(\mathbf{x})) = 1$ and correlation function $\langle \mathbf{y}(\mathbf{x_1}) \mathbf{y}(\mathbf{x_2}) \rangle =\exp\left(-\frac{|\mathbf{x}_1 - \mathbf{x}_2|^2}{2\sigma^2}\right)$ with
correlation length $\sigma$ (see Methods). The two parameters, correlation length $\sigma$ and amplitude $\eta$ were expressed in units of the lattice constant $r$. \\
\textbf{Fig.~\ref{fig_9}A,B} illustrates this procedure. RGCs are shifted in a coordinated manner across the plane, correlated in both direction and magnitude of the shift. The determinant of the Jacobian
\begin{equation}
\det J(\mathbf{x}) = \det\left(
\begin{matrix}
\frac{\partial y_1(\mathbf{x})}{\partial x_1} & \frac{\partial y_1(\mathbf{x})}{\partial x_2}\\
\frac{\partial y_2(\mathbf{x})}{\partial x_1} & \frac{\partial y_2(\mathbf{x})}{\partial x_2}\\
\end{matrix}
\right)\,,
\end{equation}
measures the local change of RGC lattice constant.  In regions of negative $\det J$, RGCs are closer together than average, in regions of positive $\det J$, RGCs are further apart (contour lines in \textbf{Fig.~\ref{fig_9}B}). In primates, regions of higher density are predicted to have higher cortical magnification and vice versa \cite{Wassle1989}. The RGC mosaics with correlated positional noise therefore imply local fluctuations in the cortical magnification factor on the scale of the noise correlation length. \\
\textbf{Fig.~\ref{fig_9}C-E} display unfiltered and smoothed model layouts obtained with spatially correlated noisy hexagonal mosaics as well as their amplitude spectra. As expected, orientation domain layouts exhibit a typical column spacing up to higher  disorder strengths, when the position disorder was correlated (compare \textbf{Fig.~\ref{fig_9}F} with \textbf{Fig.~\ref{fig_5}D}). Locally, Moir\'{e} interference leads to a roughly hexagonal layout of columns that is distorted on larger scales. Both, the orientation of the hexagons as well as column spacings change continuously across the layout. For weak disorder, the amplitude  spectrum still exhibits six peaks, indicating a globally hexagonal layout (\textbf{Fig.~\ref{fig_9}C}, right). For intermediate disorder local column spacing and direction of the hexagons varies to the extend that peaks can hardly be identified in the amplitude  spectrum of the resulting domain layout. In particular, the spatially varying local column spacing leads to a broader 
peak in the radially averaged amplitude spectrum with increasing disorder strength (\textbf{Fig.~\ref{fig_9}F}). This is in contrast to the case of uncorrelated disorder (cf. \textbf{Fig.~\ref{fig_5}D}). Note that experimental iso-orientation domain layouts exhibit a similarly broad peak in their marginal amplitude spectra \cite{Miller1994}. We quantified the pinwheel density of orientation domain layouts obtained with correlated noisy hexagonal RGCs (\textbf{Fig.~\ref{fig_9}G}) as a function of disorder correlation length and disorder strengths. Independent of the disorder correlation length, pinwheel densities plateau around $2\sqrt{3}$ for weak disorder and monotonically increase above a critical disorder strength. This critical disorder strength is higher, the larger the correlation length. Thus, model pinwheel densities are inconsistent with the individual values obtained for each species, no matter what the strength of disorder or correlation length is. \textbf{Fig.~\ref{fig_9}H} illustrates that the 
pinwheel density increases with increasing disorder strength largely because the overall measured column spacing increases, not because additional pinwheels appear in the layouts. In fact, the number of pinwheels per mm$^2$ is almost independent of either correlation length or disorder strength.
\subsection*{Pinwheel densities of iso-orientation domain layouts derived from PIPP mosaics}
\begin{figure}
\centering
\includegraphics[width=\linewidth]{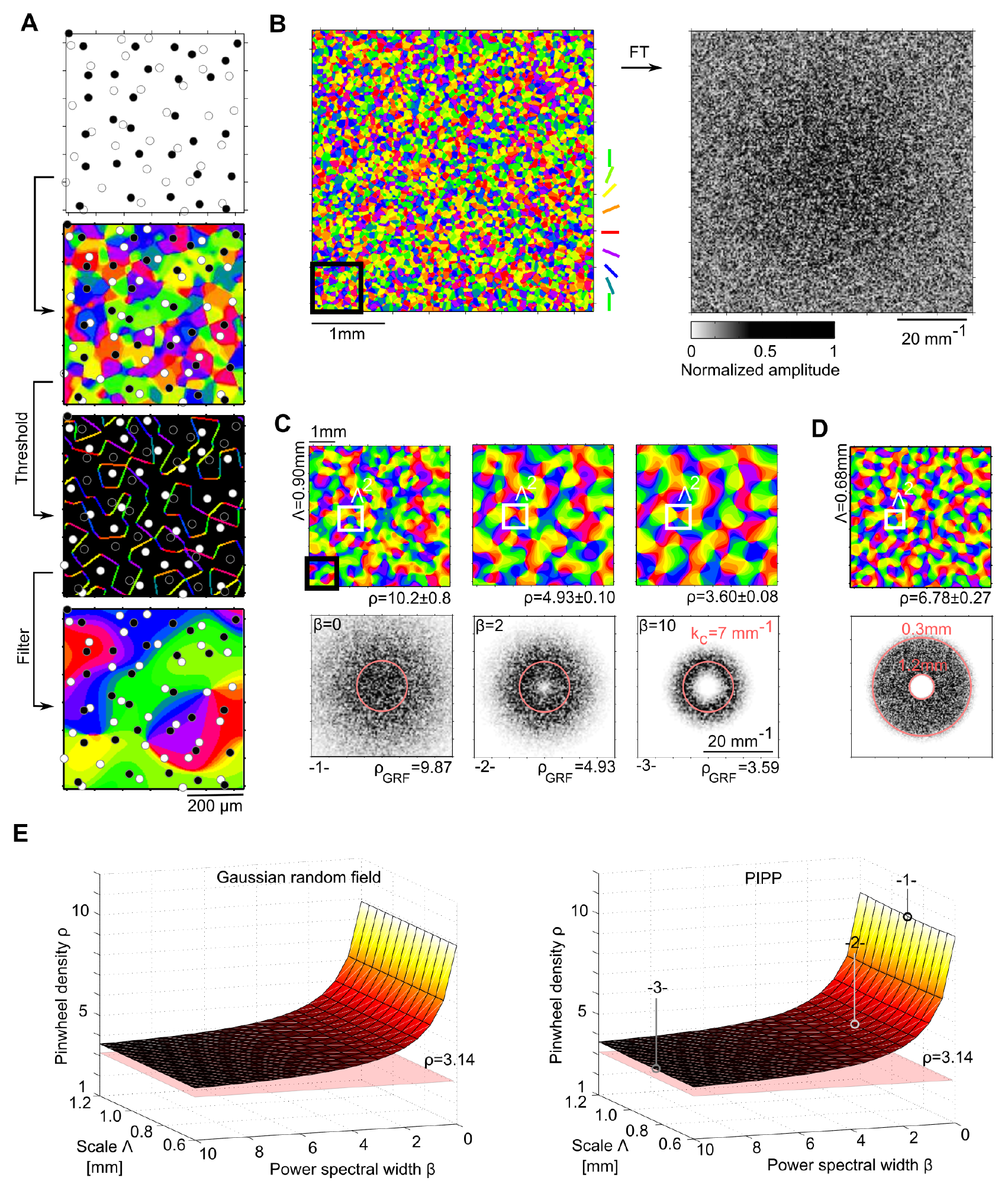}
\caption{
\textbf{Iso-orientation domain layouts obtained from PIPP RGCs with the statistical connectivity model.} 
\textbf{A} Generating orientation domain layouts from PIPP RGCs in the statistical connectivity model \cite{Ringach2004}. Top: Inset of a PIPP RGC mosaic (see Methods). Black (white) dots represent OFF (ON) cells. Middle top: unfiltered layout with RGC mosaics overlaid. Middle bottom: thresholded layout with RGC mosaics overlaid.  Bottom: thresholded and smoothed layout ($\beta = 0$) with RGC mosaics overlaid.  Scale bar indicates retinal distances, assuming PIPP parameters as in \cite{Eglen2005}.
\textbf{B} Left: larger region of the unfiltered layout shown in A (black square). Scale bar indicates retinal distances. Right: normalized amplitude spectrum of unfiltered layout shown on the left.
\textbf{C} Thresholded and smoothed layout (top) and corresponding amplitude spectrum (bottom) for filter function parameters (see Eq.~\eqref{schnabel_filter}) $\beta = 0$ (left), $\beta = 2$ (middle), and $\beta = 10$ (right). Scale bar indicates cortical distances, assuming cortical magnification factor $\approx$1, and $\Lambda = 0.9\textnormal{mm}$ (see Eq.~\eqref{schnabel_filter_k_bar} and text).
Red circles indicate $k_c = 2\pi/\Lambda$. Black square indicates inset in A, white square indicates $\Lambda^2$. 
\textbf{D} As C but filtered with Fermi band pass filters \cite{Kaschube2010}. White square (top) indicates $\Lambda = 0.68\text{mm}$, the column spacing as measured by wavelet analysis. Red circles (bottom) indicate low pass (1.2 mm) and high pass (0.3 mm) position. Pinwheel densities are stated with standard error of the mean. 
\textbf{E} Left: Analytically predicted pinwheel density of orientation domain layouts derived from Gaussian random fields \cite{Schnabel2007} as a function of filter parameter $\beta$ and spatial scale (see text). 
Right: Pinwheel density of orientation domain layouts obtained from PIPP mosaics with the statistical connectivity model as a function of filter parameter $\beta$ and spatial scale. Numbers 1-3 indicate parameter choices displayed in C.
\label{fig_10}}
\end{figure}

Finally, we examined whether the statistical connectivity model could reproduce the common design invariants with RGCs distributed in space according to a pairwise interacting point process (PIPP). The PIPP developed by Eglen et al. \cite{Eglen2005} is currently the experimentally best supported model for RGCs mosaics and was shown by several studies to generate RGC positions which accurately reproduce a variety of spatial statistics of RGC mosaics \cite{Eglen2005,Ringach2007, Hore2012, Schottdorf2014}. The PIPP model generates samples from a statistical ensemble of RGC mosaics by iteratively updating RGC positions to maximize a target joint probability density, specified by pairwise interactions between neighboring RGCs (for details see Materials \& Methods). Each PIPP mosaic represents a random realization of a regularly-spaced RGC mosaic with radially isotropic autocorrelograms \cite{Hore2012} and lacks long-range positional order. \textbf{Fig.~\ref{fig_10}A} depicts a realization of a PIPP with 
parameters choosen to reproduce cat RGC mosaics (for details see Materials \& Methods). We generated thresholded and smoothed iso-orientation domain layouts from PIPP RGC mosaics as from the ordered mosaics (\textbf{Fig.~\ref{fig_10}A,B} left). The lack of long-range positional order in ON and OFF mosaics prevents any Moir\'{e} interference between them. Thus, no typical spacing between adjacent column preferring the same orientation is set in the model layouts (\textbf{Fig.~\ref{fig_10}B}, right, see also \cite{Hore2012, Schottdorf2014}). Spectral power in these layouts is broadly distributed and monotonically decays with increasing spatial frequency. As a consequence, one needs to apply bandpass filtering to obtain orientation domain layouts that are at least qualitatively resembling the experimental data. We used a flexible band-pass filtering function $f(\mathbf{k})$ of the following form:
\begin{equation}
f(\mathbf{k}) = a |\mathbf{k}|^\beta \exp(-|\mathbf{k}|^2 b)\,,
\label{schnabel_filter}
\end{equation}
with $a,b,\beta >0$. The filter function was normalized such that 
\begin{equation}
\int \mathrm{d}^2 \mathbf{k}\, f(\mathbf{k}) = 2\pi\,.
\label{schnabel_filter_normalization}
\end{equation}
With this normalization, one can define the mean column spacing of the resulting layout via $\Lambda = 2\pi/\bar{k}$ with 
\begin{equation}
\bar{k}  = \int_0^{\infty}  \mathrm{d} k\, 2 \pi k f(\mathbf{k}) = 1\,.
\label{schnabel_filter_k_bar}
\end{equation}
By increasing the parameter $\beta$, the shape of the filter can be changed from Gaussian lowpass ($\beta = 0$, \textbf{Fig.~\ref{fig_10}C}, left) to wide bandpass ($\beta = 2$, \textbf{Fig.~\ref{fig_10}C}, middle) and to narrow bandpass ($\beta = 10$, \textbf{Fig.~\ref{fig_10}C}, right).  There is an additional degree of freedom in this filter definition, namely how the filter is scaled relative to the absolute physical units mm$^{-1}$ of the amplitude  spectrum. To scan a wide range of filter shapes and column spacings, we varied $\beta$ between 0 and 10 and choose the scaling such that $\Lambda$ varied between 0.6 mm and 1.2 mm, i.e. covering the entire range of experimentally observed mean column spacings in tree shrew, galago, cat, and ferret \cite{Kaschube2010, Keil2012}. We then measured the pinwheel densities of the resulting statistical connectivity model layouts (\textbf{Fig.~\ref{fig_10}E}, right), where pinwheel density was defined as the number of pinwheels within an area $\Lambda^2$. Pinwheel 
densities were independent of the scale $\Lambda$ and increased monotonically with increasing spectral width (decreasing $\beta$). They are in general substantially larger than the experimentally observed value of 3.14 (see also \textbf{Fig.~\ref{fig_10}C}) and outside of the single-species/common-design consistency range.\newline
Qualitatively, iso-orientation domain layouts generated with the PIPP RGC mosaics resembled those generated from Gaussian random field (GRF)  \cite{Wolf1998,Wolf2003, Schnabel2007} (\textbf{Fig.~\ref{fig_10}C}). In fact, we find that this resemblance is quantitative. \textbf{Fig.~\ref{fig_10}E} depicts the analytical prediction \cite{Schnabel2007} for the pinwheel density of orientation domains obtained with GRFs with a marginal amplitude spectrum corresponding to the filter function in Eq.~\eqref{schnabel_filter} (\textbf{Fig.~\ref{fig_10}E}, left). Pinwheel densities for GRF layouts and layouts obtained from PIPP mosaics with the statistical connectivity model are indistinguishable. For the pinwheel density to be consistent with at least the single-species consistency range ($\rho<3.42$), amplitude spectra had to be much more peaked ($\beta\geq17$) than experimentally observed \cite{Miller1994}. Finally, we filtered statistical connectivity model layouts with the Fermi-Filter function as used in \cite{
Kaschube2010, Keil2012} with cut-off wavelengths of 0.3 mm and 1.2 mm (\textbf{Fig.~\ref{fig_10}D}). Again, pinwheel densities were much larger than those observed in the experimental data and outside of the single-species and common-design consistency range.\\
In summary, iso-orientation domain layouts generated by the statistical connectivity model using PIPP RGC mosaics quantitatively resemble layouts derived from Gaussian random fields. Their statistics is distinct from the statistics of experimentally measured layouts.
\newpage
\section*{Discussion}
%
In this study, we examined whether the statistical connectivity model - a biologically plausible scheme of circuit disorder - is able to explain the common design of spatially aperiodic arrangements of orientation domains and pinwheels in the primary visual cortex. As an analytically tractable limiting case, we first considered the model with perfectly ordered hexagonal RGC mosaics (Moir\'{e} interference model). For this model we derived exact expressions for receptive fields and tuning curves as well as for unfiltered and filtered layouts. We found that unfiltered orientation domain layouts generated by Moir\'{e}-interference exhibited a fine-grained structure of subdomains with substantial and systematic variation in orientation preference on scales much smaller than the typical size of orientation domains. After smoothing, the resulting Moir\'{e} interference pattern could mathematically be expressed as the phase of a complex-valued field composed of six planar waves. The pinwheel density of this 
perfectly hexagonal pattern of orientation domains is 
$\rho=2\sqrt{3}\approx 3.46$. Next, we studied the layout of numerically obtained domain layouts derived from hexagonal mosaics that are randomly distorted by spatially uncorrelated disorder. We found that pinwheel density and pinwheel nearest neighbor statistics vary substantially with the degree of randomness. Nevertheless, there was no parameter regime in which all of the common design parameters matched experimental observations. Most prominently, the pinwheel density increased monotonically with increasing disorder strength. To examine the effect of noisy RGC mosaics more broadly, we introduced a more general class of noisy hexagonal mosaics, which allows for the inclusion of spatial correlations in RGC positional disorder. We found that, while RGC dipole patterns for such mosaics are inherently aperiodic, the model still predicts domain layouts that substantially deviate from experimentally observed pinwheel layouts. Finally, we studied the model with RGC mosaics derived from Eglen's random pairwise 
interacting point process. The resulting layouts lacked a typical spacing between neighboring orientation domains and, after bandpass filtering, pinwheel densities were inconsistent with the values observed for any of the four species investigated. 
\subsection*{Alternative random wiring models}
The statistical wiring model analyzed in the present study is only one representative of possible random wiring schemes. One could argue that alternative, perhaps more realistic, schemes might do a better job at reproducing the experimentally observed pinwheel layouts. There is good evidence that the spatial statistics of RGC mosaics is well approximated by Eglen's PIPP \cite{Eglen2005, Hore2012, Schottdorf2014}, and, hence, there is little freedom of choice at the retinal level. In contrast, at the next network layer, two main modifications or extensions of the statistical wiring model could be considered: (i) adding an additional layer to the feedforward network implementing the transformation of the retinal input structure by the lateral geniculate nucleus (LGN) (ii) choosing different probabilistic connectivity rules between the retinal/LGN layer and the primary visual cortex.  We argue that both modifications of the random wiring approach are unlikely to improve the consistency of the model with 
experimental data.\newline
Regarding (i), Martinez et al. \cite{Martinez2014} have recently tried to infer the mapping between RGC inputs and LGN relay cells using a statistical connectivity approach. In their model, ON and OFF cell types were homogeneously distributed and their polarity (ON or OFF) was inherited from the nearest retinal input. Connection probability between RGCs and LGN neurons was modeled as an isotropic Gaussian function of the relative distance between the RF centers of the presynaptic and postsynaptic partners. With this simple wiring scheme, together with similar connectivity rules for the population of inhibitory interneurons, several spatiotemporal properties of LGN RFs robustly agreed with the experimental data. In the architecture between the retina and the LGN proposed by these authors, the dipoles of ON- and OFF-center cells that characterize the retinal mosaic are transformed into small clusters of same-sign relay cells. The LGN ON-OFF dipoles occur at the boundaries of these clusters with LGN dipole 
orientations strongly correlating but not necessarily matching dipole orientations in the retina. Notably, dipole density and dipole angle correlation length in the LGN is not increased compared to the retina. The data and modeling by Martinez et al. suggest that the LGN mosaics do not systematically alter the spatial structure of RGC inputs beyond providing an additional source of dipole disorder. Additional disorder imposed by the LGN mosaics would likely add a uniform level of disorder to all spatial frequency components in the unfiltered orientation domain layouts predicted by RGC dipole structure. When hexagonal mosaics are considered, the disorder strength that has to be assumed to match the spatial distribution of RGC cells found in experiment is already rather high \cite{Ringach2004, Paik2011, Paik2012, Hore2012, Schottdorf2014}. Additional noise is likely to obstruct any remaining Moir\'{e} interference. We therefore speculate that when considering an additional LGN layer, after smoothing, domain 
layouts for both, the noisy Moir\'e interference model and the model with PIPP mosaics would be similar to those obtained with PIPP mosaics \cite{Ringach2007, Hore2012, Schottdorf2014}. As we have shown in the present study, the spatial statistics of these layouts resembles those derived from Gaussian random fields and is inconsistent with the data obtained for any of the four species analyzed (cf. \textbf{Fig.~\ref{fig_10}}, see also \cite{Kaschube2010, Wolf2003}).\newline
A similar argument can be made for alternative probabilistic connectivity rules between the retinal/LGN layer and V1. As our analysis shows, the dipole structure emerging from ``realistic" RGC mosaics (be it very noisy hexagonal mosaics or PIPP mosaics) is spatially fine-grained because dipole angles vary over short distances in cortical space relative to the typical size of an iso-orientation domain. For this reason, the statistical connectivity model requires an additional smoothing step (cf. \textbf{Figs.~\ref{fig_4},\ref{fig_5},\ref{fig_10}}) to yield smooth orientation domain layouts as observed in experiment \cite{Ohki2005, Ohki2006}. Unless the connectivity rule is assumed to specifically select dipoles with a similar angle from a larger spatial region of the retina, or neurons within an iso-orientation domain are assumed to choose one particular dipole to receive the input from and ignore all other dipoles in the vicinity, such spatial averaging within the cortical layer will always be required no 
matter what the actual probabilistic connectivity rule is. Domain layouts resulting from such spatial averaging of weakly correlated dipole angles (see also \cite{Schottdorf2014}) are likely to follow layout statistics that resemble those of Gaussian random fields, independently of the connectivity rule assumed. If neurons are assumed to select specific dipoles out of the repertoire of ``available'' ones, then the overall spatial layout of RGC dipoles is not informative about the resulting domain layout, which contradicts the main hypothesis of the statistical wiring model.\newline
Ultimately, the key experiment to provide support for the random wiring approach consists of determining both, the orientation domain layout and the retinotopic map in a single animal and, in a second step, correlate these with the spatial arrangement of RGCs in the same animal. This challenging experimental task is still awaiting its completion.
\subsection*{Spatial irregularity by disorder or optimization} 
So far, the only model class able to robustly reproduce all common design parameters, describes the formation of orientation domain layouts as a deterministic optimization process converging to quasi-periodic pinwheel-rich orientation domain patterns associated with and stabilized by a matching system of intrinsic horizontal connections \cite{Wolf2005, Kaschube2008, Kaschube2010}. Irregular layouts of orientation domains dynamically emerge as a consequence of large-scale circuit optimization of domain patterns and intrinsic circuits. Is this agreement between model and data good evidence for global circuit optimization or are there simple alternative explanations such as the random feedforward wiring hypothesis that can explain the invariant statistical properties of orientation domains? Qualitatively, it is in fact tempting to attribute the spatial irregularity and apparent randomness of pinwheel layouts in V1 to some general kind of ``biological noise". In this view, the quantitative laws of pinwheel 
organization that Kaschube et al. found \cite{Kaschube2010} would then be conceived as outcome of a largely random process underlying the emergence of orientation domains. By now, however, all proposals based on the assumption of disorder as the determinant of spatial irregularity have failed to reproduce the common design parameters and laws that have now been observed in four divergent species.\newline
Orientation domain layouts obtained from statistical ensembles of Gaussian random fields \cite{Wolf1998, Wolf2003, Schnabel2007} as well as phase randomized layouts derived from experimental data \cite{Kaschube2010}, exhibit pinwheel densities that are substantially higher than experimentally observed. Importantly, most dynamical models for the development of orientation domains produce such Gaussian random domain layouts during the initial emergence of orientation selectivity \cite{Wolf1998, Wolf2003}. Therefore approaches based on ``frozen'' early states of such models are also ruled out by the existing data (see \cite{Keil2011}).  The present study shows that a mechanistic and biologically plausible feedforward model of the early visual pathway based on (i) noisy hexagonal placement of RGCs or (ii) a more realistic semi-regular positioning of RGCs generated by the PIPP also generates layouts distinct from experimental observations. These findings illustrate that orientation domains and pinwheels positions,
 although spatially non-periodic and irregular, follow a rather distinct set of layout laws. These laws cannot easily be accounted for by a spatial irregularity or randomness in the structure of afferent projections to visual cortical neurons.\newline
A further conceivable and potentially critical source of stochasticity that is often overlooked is randomness within intracortical circuits. The mean field approach employed in modeling approaches to orientation domain layouts, such as the long-range interaction model, represents an idealization of a complex network, in which every neuron is characterized by its own set of inputs and outputs. These inputs and outputs will, at least to some extent, be stochastic. How and to what extent randomness in intracortical connections may affect and shape orientation domain layouts is currently an open question. In that respect, it is interesting to note that model networks for largely stochastic intracortical circuits are able to generate and robustly maintain orientation selective responses to afferent inputs and can lead to highly coherent orientation domains \cite{Ernst2001, Hein2014}.
\subsection*{Hexagonal order of orientation domains}
Paik \& Ringach reported indications of hexagonal order in visual cortical orientation domain patterns of tree shrew, ferret, cat, and macaque monkey \cite{Paik2011, Paik2012}. Two recent studies have casted doubt on the hypothesis that this hexagonal order echoes hexagonal or quasi-hexagonal arrangements of ganglion cell mosaics in the retina \cite{Hore2012, Schottdorf2014}. Hore et al. showed that noisy hexagonal lattices do not capture the spatial statistics of parasol cell mosaic. Moreover, the positional correlations in measured mosaics extends to only $200-350\text{ \textmu m}$, far less than required for generating Moir\'{e} interference \cite{Hore2012}. More generally, Schottdorf et al. studied the spatial arrangement of RGC dipole angles in cat beta cell and primate parasol RF mosaics \cite{Schottdorf2014}. According to the statistical wiring hypothesis, dipole angle correlations should follow the spatial correlations of preferred orientations in the primary visual cortex, i.e. positively correlate 
on short scales (0-300 \textmu m) and negatively correlated on larger scales (300-600 \textmu m) in the retina. By introducing a positive control point process that (i) reproduces both, the nearest neighbor spatial statistics and the spatial autocorrelation structure of parasol cell mosaics and (ii) exhibits a tunable degree of spatial correlations of dipole angles, they were able to show that, given the size of available data sets, the presence of even weak angular correlations in the data is very unlikely.\newline
If not from the structure of RGC mosaics, where does the apparent hexagonal organization in orientation domains come from? A variety of self-organization models on all levels of biological detail have been shown to generate orientation domains with hexagonal arrangements, e.g. \cite{Malsburg1973, Grabska-Barwinska2008, Reichl2009, Keil2011, Reichl2012a, Reichl2012b, Ernst2001} (notably including the earliest theory for the self-organization of orientation preference by von der Malsburg in 1973). Thus, hexagonal order, even if present, would not provide specific evidence in favor of Moir\'{e} interference between RGCs. Future work will have to elucidate whether the long-range interaction model for orientation domains \cite{Wolf2005, Kaschube 2008, Kaschube2010} can not only explain the common design statistics, but at the same time account for the observed hexagonal order in the visual cortex.
\subsection*{The impact of retinal orientation biases on visual cortical architecture}
Compared to the dense sampling of stimulus space by cortical neurons, the repertoire of detectors on the retina that input into a given cortical area is limited. For the cat visual pathway, Alonso et al. estimated the number of LGN X-relay cells converging onto a single simple cell in V1 to be $\sim$20-40 \cite{Alonso2001}, based on measuring the probability of finding a connection between individual geniculate and cortical neurons with overlapping receptive fields. This estimate was later confirmed by directly measuring population receptive fields of ON and OFF thalamic inputs to a single orientation column \cite{Jin2011}. With an expansion of around 1.5-2.0 from X-cells in the retina to X-relay cells in the LGN \cite{Illing1981,Peters1993}, each simple cell in V1 receives on average input from only  $\sim$10-25 RGCs. This not only implies that random afferent inputs to cortical neurons might seed groups of V1 neurons with similar orientation preferences but also that they might in fact impose substantial \
textit{biases} on the preferred orientation that can be adopted by the cortical neurons. The postnatal development of orientation columns could then be imagined as a dynamical activity-dependent process which refines and remodels an initial set of small biases provided by the RGC mosaic model through Hebbian learning rules and other mechanisms of synaptic plasticity. The up to now most striking experimental evidence that retinal organization can impose local biases in V1 function architecture was revealed by the finding that the pattern of retinal blood vessels can specifically determine the layout of ocular dominance columns in squirrel monkey \cite{Adams2002} (for a modeling study see \cite{Giacomantonio2007}).\\
Dynamical models of orientation column formation generally assume no a priori constraints or biases as to which preferred orientation a given position in the cortical surface can acquire. Usually random initial conditions determine which instance from the large intrinsic repertoire of stable potential domain layouts is adopted. Including seeds and biases derived from RGC mosaics in such models for the dynamical formation of V1 orientation domain layouts may elucidate the potentially complex interplay between a sparse set of subcortical feedforward constraints and self-organization in a dense almost continuum-like intracortical network. The present study provides a detailed description of a candidate set of such subcortical biases and, therefore, can serve as a foundation for such future investigations.
\subsection*{The common design as benchmark for models of visual cortical development and function}
The common design invariants comprise four distinct functions in addition to the apparently invariant pinwheel density. As such, they represent a rather specific quantitative characterization of orientation domain layouts. It is, thus, not surprising that entire model classes have been rejected based on whether their predictions match these invariants.\newline
Since the discovery of visual cortical functional architecture more than fifty years ago, a large number of models based on a variety of circuit mechanisms has been proposed to account for their postnatal formation (see \cite{Erwin1995, Swindale 1996, Goodhill2007} for reviews). Many of these models are explicitly or implicitly based on optimization principles and attribute a functional advantage to the intriguing spatial arrangement of orientation domains.  Because many, even mutually exclusive, models could qualitatively account for main features of orientation domain layouts such as the presence of pinwheels or the roughly periodic arrangement of columns, theory could not provide decisive evidence on 
whether to favor one hypothesis over another. It is only in recent years that the large available data sets have started to allow for a rigorous quantitative analysis of visual cortical architecture and its design principles in distinct evolutionary lineages.\newline
To date, abstract optimization models have been analyzed most comprehensively, providing in many case the complete phase diagram. For instance, Reichl et. al. systematically evaluated energy-minimization-based models for the coordinated optimization of orientation preference and ocular dominance layouts \cite{Reichl2012a, Reichl2012b}. By quantitatively comparing model solutions to the common design, they were able to rule out a whole variety of otherwise intuitive principles for their emergence. It is desirable to obtain a similarly quantitative understanding of more detailed models for the formation of orientation domains. In this regard, the analysis of abstract models is informative because there is a many-to-one relationship between detailed models of the visual cortical pathway and those abstract formulations. Abstract models often can be shown to be representative of an entire universality class and, once comprehensively characterized, the questions becomes whether more complex modeling schemes are 
simply complicated instantiations of such a class.\newline
For models of an intermediate degree of realism, semi-analytical perturbation methods can be employed to explicitly provide this mapping. Using this approach, Keil \& Wolf studied orientation domain layouts predicted by a widely used representative of a general optimization framework  \cite{Keil2011}. According to this framework, the primary visual cortex is optimized for achieving an optimal tradeoff between the representation of all combinations of local edge-like stimuli, i.e. all positions in the visual field and all orientations, and the overall continuity of this representation across the cortical surface\cite{Durbin1990}. While this framework has successfully explained a variety of qualitative aspects of orientation domain design, e.g. \cite{Durbin1990, Cimponeriu2000,Carreira2004}, the authors found quantitative disagreement with the common design in all physiologically realistic parameter regimes of the representative model \cite{Keil2011}. Their analysis enabled an unbiased comprehensive search of 
the model's parameter space for a match to the experimental data and indicated alternative more promising optimization hypotheses to explain the experimentally observed V1 functional architecture. \newline
Although the statistical wiring model is still rather simplistic, it is hard to make analytical or semi-analytical progress as soon as RGC mosaics with the necessary degree of realism are considered. In this case, the question of whether models account for the cortical architecture can only be answered with the approach we have pursued here, i.e. by systematic comparison between their solutions, experimental data, as well as predictions from minimal approaches. \newline
The results presented here show that this approach can indeed be successfully applied to rule out candidate mechanisms as sufficient explanations for the emergence of V1 functional architectures. We expect a re-examination of the quantitative predictions of other modeling approaches to be highly informative about candidate mechanisms for the formation of V1 functional architecture. \newline
This present study provides the first systematic assessment as to whether the common design of orientation domains could result from an inherently random process, as realized through the local feedforward structure of the early visual pathway rather than an optimization process coordinated on large scales. Given the disagreement between the layouts predicted by the statistical wiring model and the data, global circuit optimization as proposed by the long-range interaction model still stands as the only theory known to be capable of explaining the common design of orientation domains in the primary visual cortex. 
\section*{Acknowledgments}
We are grateful to Z. Kisv\'arday (University of Debrecen, Debrecen, Hungary) for sharing optical imaging data. We thank Matthias Kaschube (Institute for Advanced Studies \& Johann Wolfgang Goethe University, Frankfurt am Main, Germany) for providing his original PV-WAVE$^\circledR$ routines for the pinwheel statistics analysis and numerous stimulating discussions. \\
We thank Tim Gollisch, Ian Nauhaus, Kristina Nielsen, Joscha Liedtke, Conor Dempsey, and Lars Reichl for fruitful discussions and Bettina Hein, Markus Helmer and Juan Daniel Florez-Weidinger for comments on earlier versions of the manuscript. This work was supported by the HFSP (http://www.hfsp.org), BMBF (http://www.bmbf.de), DFG (http://www.dfg.de), and the MPG (http://www.mpg. de). Grant numbers SFB 889, BFL 01GQ0921, 01GQ0922, BCCN 01GQ0430, 01GQ1005B, 01GQ07113 and BFNT 01GQ0811. This work was supported in part by the National Science Foundation (http://www.nsf.gov). Grant number PHY05-51164. A Boehringer Ingelheim Fonds PhD fellowship to Manuel Schottdorf is gratefully acknowledged. The funders had no role in study design, data collection and analysis, decision to publish, or preparation of the manuscript.
%
%
%
%
%
%
%
\section*{Methods}
\subsection*{Analysis of model and experimentally obtained orientation domain layouts}
Column spacing and pinwheel statistics of both data and simulation were analyzed using the wavelet method introduced in \cite{Kaschube2002,Kaschube2003}. This method specifically takes into account that experimentally measured domain layouts often exhibit local variations in column spacing (as opposed to most model layouts) and is thus well-suited to unbiasedly compare the pinwheel layouts of model layouts and experimental data. Matlab source code for preprocessing of experimentally obtained layouts, column spacing analysis, and the analysis of pinwheel layouts can be found in the Supplemental Material, along with four example cases from ferret V1 to test the code. The full data set used in the present study is available on the neural data sharing platform http://www.g-node.org/.\\
For comparison between model orientation domain layouts and experimentally obtained layouts, both were analyzed with the exact same wavelet parameters settings. Raw difference images obtained in the experiments were Fermi bandpass filtered as described in \cite{Kaschube2010}. Filter parameters were adapted to the column spacings of the different species such that structures on the relevant scales were only weakly attenuated (see \cite{Kaschube2010}).\\
To determine the local column spacing of the layouts, we first calculated wavelet coefficients of an image $I(\mathbf{x})$, averaged over all orientations
\begin{eqnarray}
 \Psi(\mathbf{y},\Lambda) = \int\frac{\mathrm{d}\varphi}{\pi}\,\left| \int \mathrm{d}^2\mathbf{x}\, I(\mathbf{x})\cdot \phi_\mathbf{y}(\mathbf{x},\Lambda,\varphi)\right|\label{wavelettransformation}
\end{eqnarray}
where $\mathbf{y}$ is the position, $\varphi$ the orientation and $\Lambda$ the scale of the wavelet $\phi_\mathbf{y}(\mathbf{x},\Lambda,\varphi)$. We used complex-valued Morlet wavelets composed of a Gaussian envelope and a plane wave
\begin{eqnarray}
 \phi(\mathbf{x})=\frac{1}{\sigma}\exp\left(-\frac{\mathbf{x}^2}{2\sigma^2}\right) \cdot \exp(\mathfrak{i}\mathbf{k}_\phi \mathbf{x})
\end{eqnarray}
and
\begin{eqnarray}
 \phi_\mathbf{y}(\mathbf{x},\Lambda,\varphi) = \phi(\Omega^{-1}(\varphi)(\mathbf{y}-\mathbf{x})).
\end{eqnarray}
The matrix $\Omega(\varphi)$ is the two dimensional rotation matrix (Eq.~\eqref{rot}). To compute the wavelet orientation average in Eq.~\eqref{wavelettransformation}, 16 equally spaced wavelet orientations were used. For a given $\Lambda$, the parameters of the Morlet wavelet were chosen as
\begin{eqnarray}
 \mathbf{k}_\phi&=&\frac{2\pi}{\Lambda}\begin{pmatrix} 1\\0 \end{pmatrix}\\
 \sigma&=&\frac{\xi\Lambda}{2\pi}.
\end{eqnarray}
$\xi$ determines the size of the wavelet and was chosen to be $\xi = 7$, as in \cite{Kaschube2010}. This captures column spacings variations on scales larger than 4 hypercolumns while at the same time enabling robust column spacing estimation. To obtain the map of local column spacing $\Lambda_\text{local}(\mathbf{y})$, we calculated the scale $\Lambda$ with the largest wavelet coefficient
\begin{eqnarray}
\Lambda_\text{local}(\mathbf{y})=\text{argmax}_\Lambda\left( \Psi(\mathbf{y},\Lambda) \right)
\end{eqnarray}
for every position $\mathbf{y}$. \\
To estimate the pinwheel density and other pinwheel layout parameters, we used a fully automated procedure proposed in \cite{Kaschube2010}. We refer to the Supplemental Material accompanying \cite{Kaschube2010} for further details.
\subsection*{A mathematical treatment of the Moir\'{e}-Interference model}
Here, we examine the analytically most tractable variant of the statistical wiring model, in which ON and OFF center cells are localized on perfectly hexagonal lattices $\mathcal{L}$, (\textbf{Fig.~\ref{fig_1}B}), that may exhibit different lattice constants $r$ and $r'$,
\begin{eqnarray}
 \mathcal{L}=\left(\begin{pmatrix} 1\\0 \end{pmatrix}k+\frac{1}{2}\begin{pmatrix} 1\\ \sqrt{3} \end{pmatrix}l\right)f\qquad\forall\, k,l \in \mathbb{Z},
 \label{lattice_def}
\end{eqnarray}
where $f=r,r'$ is the lattice constant. Describing a rotation of the lattice vectors by the rotation matrix
\begin{eqnarray}
 \Omega(\alpha)=\begin{pmatrix} \cos(\alpha) & -\sin(\alpha)\\ \sin(\alpha) & \cos(\alpha)\end{pmatrix}\label{rot},
\end{eqnarray}
the ON mosaic is rotated by an angle $\alpha$, the OFF lattice by an angle $\alpha'$ (\textbf{Fig.~\ref{fig_1}B}). Paik \& Ringach found in numerical simulations that in this case, Moir\'{e} interference between two hexagonal RGC mosaics results in a hexagonal layouts of orientation domains \cite{Ringach2004, Paik2011, Paik2012}. We now first derive an explicit expression for cortical receptive fields $\text{RF}_\mathbf{y}$ spatially varying with $\mathbf{y}$ predicted by the model. Calculating the preferred orientation of these receptive fields then provides an explicit expression for the hexagonal domain layouts.\\
The sum in Eq.~\eqref{summation} can be evaluated analytically for \textit{rectangular} lattices using Jacobi Theta functions \cite{Weisstein}. To solve the Moir\'{e} interference model, we used the fact that every hexagonal lattice can be written as sum $\mathcal{L}=\mathcal{L}_1+\mathcal{L}_2$ of two rectangular lattices with orthogonal base vectors by separating even and odd numbers in $l$ and shifting the $l$-sum so that the x-component is equal to zero. The two rectangular lattices are
\begin{eqnarray}
 \mathcal{L}_1&=&\left(\begin{pmatrix} 1\\0 \end{pmatrix}k+\begin{pmatrix} 0\\ \sqrt{3} \end{pmatrix}l\right)f\qquad \forall \, k,l \in \mathbb{Z},\nonumber \\
\mathcal{L}_2&=&\left(\begin{pmatrix} 1\\0 \end{pmatrix}k+\begin{pmatrix} 0\\ \sqrt{3} \end{pmatrix}l+\frac{1}{2}\begin{pmatrix} 1\\ \sqrt{3} \end{pmatrix}\right)f\qquad \forall \,  k,l \in \mathbb{Z}.
\label{rectangular_lattices}
\end{eqnarray}
Evaluating the infinite Gaussian sum (Eq.~\eqref{summation}) yields the result for a single sub lattice (either ON or OFF)
\begin{eqnarray}
\text{RF}_{\alpha,r,\mathbf{y}}^\text{ON/OFF}(\mathbf{x})&=&T\left(\Theta_3\left(\mathbf{b}\,\mathbf{e}_\phi,\tau\right) \Theta_3\left(\mathbf{b}\,\mathbf{e}_r,\kappa\right)+\Theta_4\left(\mathbf{b}\,\mathbf{e}_\phi,\tau\right) \Theta_4\left(\mathbf{b}\,\mathbf{e}_r,\kappa\right)\right)\label{rfsummed}\qquad
\end{eqnarray}
where
\begin{eqnarray}
\mathbf{b}&=&\frac{\mathbf{x}\sigma_s^2+\mathbf{y}\sigma_r^2}{\sigma_s^2+\sigma_r^2}\\
 \mathbf{e}_r&=&-\frac{\pi}{r}\begin{pmatrix} \cos(\alpha)\\ \sin(\alpha) \end{pmatrix}\\
\mathbf{e}_\phi&=&-\frac{\pi}{\sqrt{3} r}\begin{pmatrix} -\sin(\alpha)\\\cos(\alpha) \end{pmatrix}\\
 \tau&=&\text{e}^{-\frac{2 \pi ^2 \sigma_r^2\sigma_s^2}{3 r^2(\sigma_r^2+\sigma_s^2)}}\\
\kappa&=&\text{e}^{-\frac{2 \pi ^2 \sigma_r^2\sigma_s^2}{r^2(\sigma_r^2+\sigma_s^2)}}\\
T&=&\frac{2\pi\sigma_r^2\sigma_s^2}{\sqrt{3}r^2(\sigma_r^2+\sigma_s^2)}\exp\left(-\frac{(\mathbf{x}-\mathbf{y})^2}{2(\sigma_s^2+\sigma_r^2)}\right)
\end{eqnarray}
and $\Theta_3$ and $\Theta_4$ are the third and fourth Jacobi theta functions \cite{Weisstein}. The cortical receptive fields $RF_\mathbf{y}(\mathbf{x})$ are obtained by summing the ON and OFF sublattices 
\begin{eqnarray}
 \text{RF}_\mathbf{y}(\mathbf{x})=\text{RF}_{\alpha,r,\mathbf{y}}^\text{ON}(\mathbf{x})-\text{RF}_{\alpha',r',\mathbf{y}}^\text{OFF}(\mathbf{x})\,,
\label{eqnrf}
\end{eqnarray}
where $\alpha$ ($\alpha'$) and $r$ ($r'$) are the angle and the lattice spacing of the ON (OFF) lattice. The inset of a receptive field in \textbf{Fig.~\ref{fig_3}B} shows a plot of Eq.~\eqref{eqnrf}. These receptive fields resemble simple cell receptive fields in V1 with a size of about $1^\circ$ \cite{Jones1987, Mazer2002}. An implementation/visualization of the equations for receptive fields and their amplitude spectra can be found in the Supplemental Material.
\subsection*{Tuning curves from receptive fields}
The response $R_\mathbf{y}$ of a neuron with receptive field $\text{RF}(\mathbf{x})$ to a sine wave grating can be calculated using $\text{L}(\mathbf{x})=\exp(-\mathfrak{i}\mathbf{k}\mathbf{x})$ as a stimulus in Eq.~\eqref{response}. Evaluating the integral then corresponds to Fourier transforming the receptive field $\text{RF}(\mathbf{x})$. Denoting the Fourier transform of the receptive field as
\begin{eqnarray}
 \mathcal{R}_\mathbf{y}(\mathbf{k})=\frac{1}{2\pi}\int \mathrm{d}^2\mathbf{x}\, \text{RF}_\mathbf{y}(\mathbf{x})\text{e}^{- \mathfrak{i}\mathbf{k}\mathbf{x}},
\end{eqnarray}
we refer to the absolute value $|\mathcal{R}_\mathbf{y}(\mathbf{k})|$ as the amplitude spectrum of the receptive field. Given the above definition, the amplitude spectrum represents the response to a sine wave grating with wave vector $\mathbf{k}=(k \cos(\vartheta), k \cos(\vartheta))$, where $\vartheta$ is the grating orientation and $k$ its spatial frequency. A \textit{tuning curve} for spatial frequencies $k$ and orientations $\vartheta$ is given by
\begin{eqnarray}
 \text{TC}(\vartheta,k) = |\mathcal{R}_\mathbf{y}(k\cos(\vartheta),k\sin(\vartheta))|\label{tcdef}.
\end{eqnarray}
We calculated the Fourier transform of Eq.~\eqref{eqnrf} by transforming Eq.~\eqref{rgcrf} and subsequently summing over the two rectangular lattices $\mathcal{L}_1$ and $\mathcal{L}_2$ in Eq.~\eqref{rectangular_lattices}. Interchanging summation and integration is valid because all infinite sums are uniformly convergent. The result is
\begin{eqnarray}
 \mathcal{R}(\mathbf{k})_{\alpha,r,\mathbf{y}}^\text{ON/OFF}=U\left(\Theta_3(\mathbf{c}\mathbf{e}_\phi,\nu)\Theta_3(\mathbf{c}\mathbf{e}_r,\zeta)+\Theta_4(\mathbf{c}\mathbf{e}_\phi,\nu)\Theta_4(\mathbf{c}\mathbf{e}_r,\zeta)\right)\,,
 \label{rpart}
\end{eqnarray}
where
\begin{eqnarray}
U&=&\frac{2\pi \sigma_r^2\sigma_s^2}{\sqrt{3}r^2}\exp\left(-\mathfrak{i}\mathbf{k}\mathbf{y}-\frac{1}{2}\mathbf{k}^2\left(\sigma_s^2+\sigma_r^2\right)\right)\\
 \mathbf{c}&=&(\mathbf{y}-i\sigma_s^2\mathbf{k})\\
\nu&=&\exp\left(\frac{-2\pi^2\sigma_s^2}{3 r^2}\right)\\
\zeta&=&\exp\left(\frac{-2\pi^2\sigma_s^2}{r^2}\right).
\end{eqnarray}
The Fourier transform of cortical receptive fields is given by the sum of the ON and OFF sublattice Fourier transforms
\begin{eqnarray}
 \mathcal{R}_\mathbf{y}(\mathbf{k})= \mathcal{R}(\mathbf{k})_{\alpha,r,\mathbf{y}}^\text{ON}- \mathcal{R}
 (\mathbf{k})_{\alpha',r',\mathbf{y}}^\text{OFF}\label{specfu}\,,
\end{eqnarray}
where we suppressed the dependencies on $\alpha, \alpha', r, r'$ on the left hand side. Receptive fields depend on the two scales $\sigma_r$ and $\sigma_s$. With increasing $\sigma_s$, more RGCs are pooled to form the cortical receptive field. If the cortical receptive field is dominated by more than two RGCs, it can exhibit multiple ON and OFF subregions. The spatial arrangement of these ON and OFF subregion mirrors the hexagonal lattices of the ON and OFF center RGCs. The parameter $\sigma_s$ must be of a minimal size since for very small values of many cortical cells are connected with only a single, dominant RGC input and exhibit no orientation selectivity. Varying $\sigma_r$ does not qualitatively change the shape of cortical receptive fields.
\subsection*{Extracting preferred orientation and spatial frequency from amplitude spectra of receptive fields}
\begin{figure}
\begin{center}
\includegraphics[width=\linewidth]{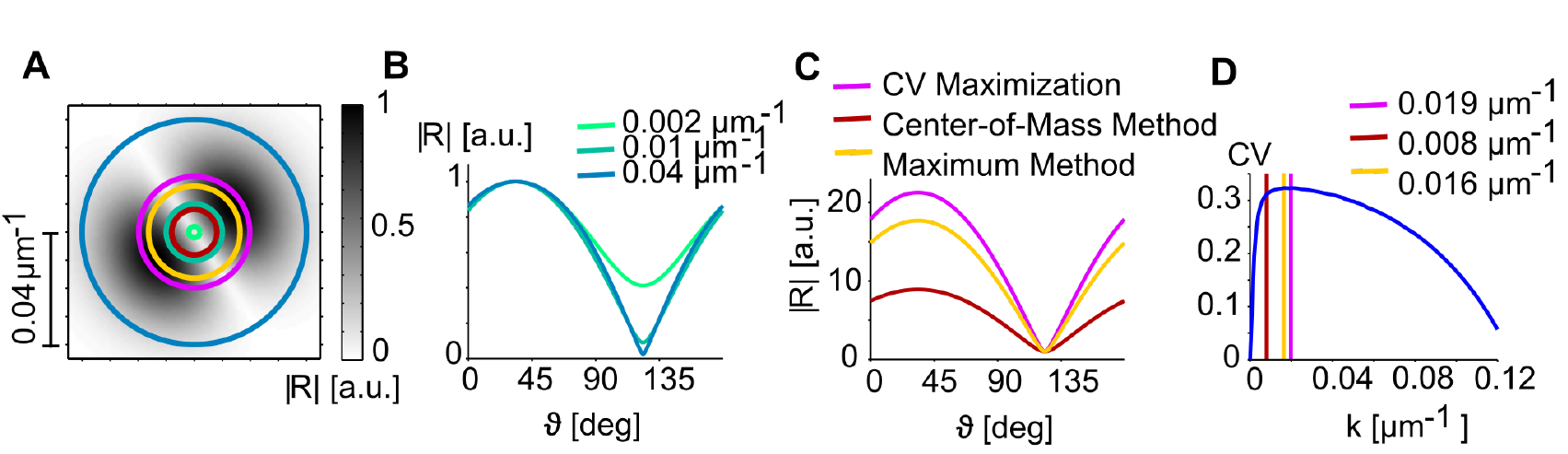}
\end{center}
\caption{\textbf{A comparison of different methods for extracting RF parameters from their amplitude spectrum} \textbf{A} Amplitude spectrum $|R(\mathbf{k})|$ of a simple cell receptive field calculated with Eq.~\eqref{specfu} in the manuscript.  Circles indicate the preferred spatial frequency as extracted by maximizing the CV (yellow), by the Center-of-Mass Method (brown), and the Maximum Method (pink). 
\textbf{B} The Tuning curves corresponding to the green, cyan and blue circles in A, normalized relative to their maximum. 
\textbf{C} The Tuning curves corresponding to the brown, the yellow and the pink circles in A, normalized relative to their minimum. 
\textbf{D} Circular variance of the various tuning curves calculated via Eq.~\eqref{cvdef} as function of spatial frequency. The brown, the yellow and the pink line correspond to the preferred spatial frequency as extracted by the three methods.}
\label{Suppl_Fig_1}
\end{figure}

For simple cell receptive fields with one ON and one OFF subregions, the amplitude spectrum $|\mathcal{R}(\mathbf{k})|$ will typically look as in \textbf{Fig.~\ref{Suppl_Fig_1}A}. We follow \cite{Ringach2007, Paik2011} and define the preferred angle as $\vartheta_\text{pref}:= \arg(\boldsymbol \mu)/2$, where 
\begin{equation}
\boldsymbol \mu=\frac{\int  \mathrm{d}^2 \mathbf{k}\, |\mathcal{R}(\mathbf{k})|\cdot \text{e}^{2\mathfrak{i}\,\text{arg}(\mathbf{k})}\mathbf{k}}{\int  \mathrm{d}^2\mathbf{k}\, |\mathcal{R}(\mathbf{k})|}\,.
\end{equation}
While there is consensus about the definition of the preferred orientation, methods for extracting the preferred spatial frequency differ within the literature. Ringach proposed to use $k_\text{pref} = |\boldsymbol \mu|$ \cite{Ringach2007}, referred to as \textit{Center-of-Mass Method}. 
More commonly, the circular variance $CV(k)$ \cite{Worgotter1987,Swindale1998,Bredfeldt2002}
\begin{eqnarray}
 \text{CV}(k)=\frac{\left|\int_0^{2\pi}  \mathrm{d}\vartheta\, \text{TC}(\vartheta,k)\text{e}^{2\mathfrak{i}\vartheta}\right|}{\int_0^{2\pi}  \mathrm{d}\vartheta\, \text{TC}(\vartheta,k)},
\label{cvdef}
\end{eqnarray}
is first computed as a measure of orientation selectivity at a given spatial frequency $k$. Maximizing the circular variance across all spatial frequencies is then performed to obtain an estimate of preferred spatial frequency:
\begin{eqnarray}
 k_\text{pref}=\underset{\{k\}}{\text{argmax}}\left(\text{CV}(k)\right).
\end{eqnarray}
We refer to this method as \textit{CV maximization}. Finally, one can use the maximum of the amplitude spectrum
\begin{eqnarray}
 k_\text{pref}=\underset{\{k\}}{\text{argmax}} \left(|\mathcal{R}(\mathbf{k})|\right),
\end{eqnarray}
as an estimate of the preferred spatial frequency (\textit{Maximum method}). We argue that Maximum method and CV maximization in most cases yield  similar results. They extract preferred spatial frequencies that one would obtain when searching for the ``strongest response'' by presenting a set of gratings of varying orientation and spatial frequency to a subject \cite{Bredfeldt2002, Issa2000, Movshon1978}. In contrast, estimates made with the center-of-mass method are usually substantially smaller then this intuitive measure (\textbf{Fig.~\ref{Suppl_Fig_1}C,D}). As a consequence, the orientation selectivity index defined as 
\begin{eqnarray}
 \text{OSI}=\frac{\left|\int_0^{2\pi}  \mathrm{d}\vartheta\, \text{TC}(\vartheta,k_\text{pref})\text{e}^{2\mathfrak{i}\vartheta}\right|}{\int_0^{2\pi}  \mathrm{d}\vartheta\, \text{TC}(\vartheta,k_\text{pref})}\,
\label{osidef}
\end{eqnarray}
for all three methods, will usually be substantially smaller, when estimated with the Center-of-Mass method compared to the other two methods. Among all three methods, the Maximum method has the advantage that its estimates are unaffected by monotonic nonlinearities applied to $R$ commonly used to convert it to a firing rate of a neuron. For this reason, the Maximum method is our method of choice for extracting the preferred spatial frequency from amplitude spectra of receptive fields.
\subsection*{Extracting the spatial progression of preferred orientation}
Using the equations for the receptive fields of cortical neurons in the Moir\'e interference model, we extracted the spatial progression preferred orientation and spatial frequency from their squared amplitude spectrum:
\begin{equation}
\begin{split}
|\mathcal{R}_{\mathbf{y}}(\mathbf{k})|^2\propto&\exp\left(-\mathbf{k}^2(\sigma_s^2+\sigma_r^2)\right)\cdot\left|\underbrace{\Theta^{\alpha,r}_3\Theta^{\alpha,r}_3+\Theta^{\alpha,r}_4\Theta^{\alpha,r}_4-\Theta^{\alpha',r'}_3\Theta^{\alpha',r'}_3-\Theta^{\alpha',r'}_4\Theta^{\alpha',r'}_4}_{G_{\mathbf{y}}(\mathbf{k})}\right|^2\,,
\end{split}
\end{equation}
with abbreviation $\Theta^{\alpha,r}_i\Theta^{\alpha,r}_i = \Theta_i(\mathbf{c}\mathbf{e}_\phi(\alpha,r),\tau)\Theta_i(\mathbf{c}\mathbf{e}_r(\alpha,r),\zeta)$. $|\mathcal{R}(\mathbf{k})|^2$ is composed of a rotationally symmetric Gaussian envelope and a non-rotationally symmetric part $G_{\mathbf{y}}(\mathbf{k})$ varying in space $\mathbf{y}$. To calculate the preferred orientation $\vartheta_\text{pref}$ and spatial frequency $k_\text{pref}$, we expanded the non-rotationally symmetric part $|G_{\mathbf{y}}((\mathbf{k})|^2$ to quadratic order in $\mathbf{k}$:
\begin{eqnarray}
 |\mathcal{R}_{\mathbf{y}}(\mathbf{k})|^2\approx\exp\left(-\mathbf{k}^2(\sigma_s^2+\sigma_r^2)\right)\left(|G_{\mathbf{y}}^0|^2+\frac{1}{2}\begin{pmatrix} k_1&k_2 \end{pmatrix}\mathcal{H}_{\mathbf{y}}\begin{pmatrix} k_1\\k_2 \end{pmatrix}\right)\,,\label{quadapp}
\end{eqnarray}
where $\mathcal{H}_{\mathbf{y}}$ is the Hessian matrix 
\begin{eqnarray}
\mathcal{H}_\mathbf{y}=\left.\begin{pmatrix} \frac{\partial^2 |G_{\mathbf{y}}|^2}{\partial k_1^2} & \frac{\partial^2 |G_{\mathbf{y}}|^2}{\partial k_1\,\partial k_2}\\\frac{\partial^2 |G_{\mathbf{y}}|^2}{\partial k_2\,\partial k_1} & \frac{\partial^2 |G_{\mathbf{y}}|^2}{\partial k_2^2} \end{pmatrix}\right|_{\mathbf{k}=\mathbf{0}}\equiv\begin{pmatrix} a_\mathbf{y} & b_\mathbf{y}\\b_\mathbf{y} & c_\mathbf{y} \end{pmatrix}
\end{eqnarray}
and $G_{\mathbf{y}}^0=G_{\mathbf{y}}(\mathbf{k}=0)$. Since $G_{\mathbf{y}}(\mathbf{k}) = G_{\mathbf{y}}(-\mathbf{k})$, this Taylor expansion only contains terms of even power in $\mathbf{k}$. \newline
Using the fact that
\begin{eqnarray}
h(\theta)=\begin{pmatrix}\cos\theta &\sin\theta\end{pmatrix}\mathcal{H}_\mathbf{y}
\begin{pmatrix}\cos\theta\\ \sin\theta\end{pmatrix}=a_\mathbf{y} \cos^2\theta+2b_\mathbf{y} \cos\theta\sin\theta+c_\mathbf{y}\sin^2\theta
\end{eqnarray}
yields the second directional derivative in the direction of $(\cos\theta,\sin\theta)$, $\vartheta_\text{pref}$ can be found as the maximum of $h(\theta)$, i.e. the direction of largest increase in amplitude spectrum,
\begin{eqnarray}
 \vartheta_\text{pref}(\mathbf{y})=\text{atan}\left(\frac{\sqrt{(a_\mathbf{y}-c_\mathbf{y})^2+4b_\mathbf{y}^2}-a_\mathbf{y}+c_\mathbf{y}}{2b_\mathbf{y}}\right)
 \label{fullopm}.
\end{eqnarray}
This formula for $\vartheta_\text{pref}(\mathbf{y})$ represents an expression for the orientation domain layouts produced by the Moir\'e interference model for hexagonal RGC lattices.
Finally, the preferred spatial frequency was defined as the maximum of the approximated $|\mathcal{R}_\mathbf{y}(\mathbf{k})|^2$ function in the direction of $\vartheta_\text{pref}(\mathbf{y})$ (see section above).\\
To calculate the smoothed domain layout of the Moir\'e interference model analytically, we identified the low frequency components of our analytical solution. To this end, we expanded the Jacobi theta functions in Eq.~\eqref{specfu} \cite{Weisstein} 
\begin{eqnarray}
 |\mathcal{R}_\mathbf{y}(\mathbf{k})|^2=\sum_j \left(C^j_\mathbf{y}\exp\left(-\frac{1}{2\sigma^2}(\mathbf{k}-\mathbf{a}^j_\mathbf{y})^2\right)+C^j_\mathbf{y}\exp\left(-\frac{1}{2\sigma^2}(\mathbf{k}+\mathbf{a}^j_\mathbf{y})^2\right)\right)
\label{gaufo}
\end{eqnarray}
where $C^j_\mathbf{y}$ and $\mathbf{a}^j_\mathbf{y}$ are determined by Eq.~\eqref{specfu}. According to this equation, the power spectra of receptive fields in the Moir\'e interference model is represented by an infinite sum of Gaussians, each mirrored at the origin $(0,0)$ of Fourier space. The preferred orientation of a receptive field represented by such an infinite sum is set by the direction in which the "center-of-mass" of the Gaussians is located. Due to the symmetry of the power under spatial inversion, there are two peaks located at $\vartheta(\mathbf{y})$ and $\pi+\vartheta(\mathbf{y})$. The direction towards the center-of-mass of the peak is obtained through the complex number 
\begin{eqnarray}
 \mu(\mathbf{y})=\sum_j C^j_\mathbf{y}\cdot|\mathbf{a}^j_\mathbf{y}|\exp(2 \mathfrak{i}\,\text{arg}(\mathbf{a}^j_\mathbf{y}))\label{wfu}
\end{eqnarray}
with $C^j_\mathbf{y}$ and $\mathbf{a}^j_\mathbf{y}$ defined as in Eq.~\eqref{gaufo}. The preferred orientation then is $\arg(\mu(\mathbf{y}))/2$. Rewriting this sum and substituting the respective expressions for $C^j_\mathbf{y}$ and $a^j_\mathbf{y}$, we obtained
\begin{eqnarray}
 \mu(\mathbf{y})=\sum_{m,n,o,p}f(m,n,o,p)\exp(2 \mathfrak{i}\mathbf{y}(n\mathbf{e}_\phi+m\mathbf{e}_r-o\mathbf{e}_r'-p\mathbf{e}_\phi'))\label{ser},
\end{eqnarray}
with coefficients $f(m,n,o,p)$. This is a decomposition of the orientation domain layout of the Moir\'e interference model into Fourier modes, indexed by four numbers $m,n,o,p = 0,\pm1,\pm2,\dots$. \textbf{Table~\ref{mapcontrib}} lists the first terms of this series in ascending spatial frequency order. By rewriting $\mu\to z$ and selecting only the lowest contributing spatial frequencies, we obtain Eq.~\eqref{compfield}. The phase factor
\begin{eqnarray}
u_0&=&\frac{\text{e}^{\mathfrak{i} (\alpha+\alpha')} \left(\text{e}^{\mathfrak{i} \alpha'} r+\text{e}^{\mathfrak{i} \alpha} r'\right)}{\text{e}^{\mathfrak{i} \alpha} r+\text{e}^{\mathfrak{i} \alpha'} r'}
\label{constant_phase_factor}
\end{eqnarray}
is associated with an overall shift of all preferred orientations. Note that $|u_0|^2=1$. 
\begin{table}[t]
\centering
\begin{tabular}{|c|c|c|c|c|c|c|}
\hline
  m & n & o & p & $\mathbf{k}_i$ & $|\mathbf{k}_i|$ & Phase factor $u_i$ \\
  \hline
  1 & 1 & 1 & 1 & $2((\mathbf{e}_r-\mathbf{e}_r')+(\mathbf{e}_\phi-\mathbf{e}_\phi'))$ & $k_c$ & $u_0 e^{\mathfrak{i} 4\pi/3 }$  \\
 -1 &-1 &-1 &-1 & $-2((\mathbf{e}_r-\mathbf{e}_r')+(\mathbf{e}_\phi-\mathbf{e}_\phi'))$ & $k_c$ & $u_0 e^{\mathfrak{i} 4\pi/3 }$  \\
  1 &-1 & 1 &-1 & $2((\mathbf{e}_r-\mathbf{e}_r')-(\mathbf{e}_\phi-\mathbf{e}_\phi'))$ & $k_c$ & $u_0 e^{\mathfrak{i} 2\pi/3 }$  \\
 -1 & 1 &-1 & 1 & $-2((\mathbf{e}_r-\mathbf{e}_r')-(\mathbf{e}_\phi-\mathbf{e}_\phi'))$ & $k_c$ & $u_0 e^{\mathfrak{i} 2\pi/3 }$  \\
  0 & 2 & 0 & 2 & $4(\mathbf{e}_\phi-\mathbf{e}_\phi')$ & $k_c$ & $u_0$  \\
  0 &-2 & 0 &-2 & $-4(\mathbf{e}_\phi-\mathbf{e}_\phi')$ & $k_c$ & $u_0$ \\
  2 & 0 & 2 & 0 & $4(\mathbf{e}_r-\mathbf{e}_r')$ & $\sqrt{3}k_c$ & n.d.  \\
 -2 & 0 &-2 & 0 & $-4(\mathbf{e}_r-\mathbf{e}_r')$ & $\sqrt{3}k_c$ & n.d.  \\
  \hline
\end{tabular}\\
\caption{\textbf{Lowest frequency contributions of the orientation domain layout predicted by the Moir\'e interference model.} The vectors $\mathbf{k}_i$ are the wave vectors and $|\mathbf{k}_i|$ their absolute values. $u_i$ denotes the phase factors of the complex-valued coefficient $f(m,n,o,p)$ in Eq.~\eqref{ser} (see also Eq.~\eqref{compfield}). The constant phase factor $u_0$ is given by Eq.~\eqref{constant_phase_factor}. n.d. means not determined. \label{mapcontrib}}
\end{table}
\subsection*{Correlated noise on hexagonal RGC mosaics}
Realization of a random distortion field $\mathbf{y}(\mathbf{x})$ were generated by finding a complex-valued field $z(\mathbf{x})$ of which real and imaginary part correspond to dislocations in x and y direction, respectively. We constructed such a field using established methods (e.g. \cite{Schnabel2007}) in the Fourier domain. In short, we drew complex-valued amplitudes $a(\mathbf{k})$ from a Gaussian distribution satisfying $\langle a(\mathbf{k})a(\mathbf{k}')\rangle = \tilde{f}(\mathbf{k})\cdot\delta_{\mathbf{k},\mathbf{k}'}$, where $\tilde{f}(\mathbf{k})$ was a chosen power spectrum, in our case a Gaussian with width $1/\sigma$, $\sigma$ being the desired correlation length. The corresponding amplitude spectrum was then inversely Fourier transformed to obtain a complex-valued field $z(\mathbf{x})$. Real and imaginary part of this field constitute two independent real-valued Gaussian random fields, both with the desired spatial statistics. We then transformed the coordinates of the hexagonal ON/OFF 
lattice points $\mathbf{r}_i 
= (x_i, y_i)$ according to $x_i\to x_i+\eta\,\Re(z(\mathbf{r}_i))$ and $y_i\to y_i+\eta\,\Im(z(\mathbf{r}_i))$. For the displacements of ON and OFF lattices, we used two independent complex-valued Gaussian random field realizations. Source code to generate hexagonal RGC mosaics with correlated spatial noise along with Matlab code for visualization is part of the supplementary material to this manuscript.
%
\subsection*{Generating PIPP RGCs mosaics}
We generated RGC mosaics with a pairwise interacting point process using the code published by Schottdorf et al. \cite{Schottdorf2014} derived from the method developed in \cite{Eglen2005, Ringach2007}.
\clearpage
\section*{Supplementary Information Legends}
\subsection*{Source code S1 - Schottdorf\_Keil\_Suppl\_material.zip}
\textbf{Source code for pinwheel statistics analysis and simulating statistical connectivity model layouts with a variety of RGC mosaics}
\section*{Numerical implementation of the statistical wiring model}
We provide all necessary code to calculate single neuron properties and orientation domain layouts along with a Mathematica\textsuperscript{\textregistered} program which contains the analytical solution for both, an example single neuron and the domain layout obtained from a perfect and infinite lattice. The C-program 'calculate\_single\_neuron.cpp' calculates the same for a single neuron numerically. The C-program is provided to illustrate the use of the rfanalyzer class. The c-program 'calculate\_map.cpp' calculates the same properties as 'calculate\_single\_neuron.cpp' but for a whole array of cells. After finishing a run, this program generates a set of ascii files in which the output is stored. These files are read in and analyzed by the Matlab\textsuperscript{\textregistered} program 'plot\_results.m'. It calculates the pinwheel density, pinwheel distance distributions, mean pinwheel distance and pinwheel density fluctuations as a function of subregion size.\newline
We compiled the code with gcc [g++ (Ubuntu 4.8.2-19ubuntu1) 4.8.2] and the gsl:\newline\newline
g++ ./calculate\_single\_neuron.cpp -lgsl -lgslcblas -O3 -march=native\newline
g++ ./calculate\_map.cpp -lgsl -lgslcblas -O3 -march=native\newline\newline
For this article, we have calculated orientation domain layouts with aspect ratio 22x22$\Lambda$, sampled with 4096x4096 pixels. This corresponds to $\approx6.5\text{ \textmu m}$ per cortical unit for our standard combination of parameters ($r = r' = \text{170 \textmu m}$ and $\Delta \alpha = 7^{\circ}$).
\section*{Experimental data}
The folder 'map\_data' contains a data folder with single condition layouts and various ROIs for four ferrets cases. It also contains two Matlab\textsuperscript{\textregistered} files,'run\_analysis.m' and 'plot\_results.m' to run the analysis and display the results. The full data set used in the present study is available on the neural data sharing platform http://www.g-node.org/.
\clearpage
\printbibliography

\end{document}